\documentclass[twocolumn,showpacs,amsmath,nofootinbib,superscriptaddress]{revtex4}

\usepackage[dvips]{graphicx}
\usepackage{epsfig}
\usepackage{placeins}
\usepackage{xspace}
\usepackage[usenames,dvipsnames]{color}
\usepackage{tikz}
\usepackage{comment}
\usepackage{cases}
\usepackage{filecontents}
\usepackage{algpseudocode}
\usepackage{url}

\def\Xint#1{\mathchoice
{\XXint\displaystyle\textstyle{#1}}%
{\XXint\textstyle\scriptstyle{#1}}%
{\XXint\scriptstyle\scriptscriptstyle{#1}}%
{\XXint\scriptscriptstyle\scriptscriptstyle{#1}}%
\!\int}
\def\XXint#1#2#3{{\setbox0=\hbox{$#1{#2#3}{\int}$ }
\vcenter{\hbox{$#2#3$ }}\kern-.6\wd0}}

\def\dashint{\Xint-}

\newcommand{\imag}{\imath}

\newcommand{\EC}{\mathcal{E}}

\newcommand{\GC}{\mathcal{G}}
\newcommand{\IC}{\mathcal{I}}
\newcommand{\KC}{\mathcal{K}}
\newcommand{\NC}{\mathcal{N}}
\newcommand{\OC}{\mathcal{O}}
\newcommand{\RC}{\mathcal{R}}
\newcommand{\TC}{\mathcal{T}}
\newcommand{\cross}{{\mathsf{X}}}
\newcommand{\ccross}{{\cross\cross}}

\newcommand{\ave}[1]{\left\langle#1 \right\rangle}
\newcommand{\spave}[1]{\overline{#1}}

\usepackage{dsfont}
\newcommand{\gpset}[1]{\mathds{#1}}

\newcommand{\Nset}{\gpset{N}}

\newcommand{\plaind}{\mathrm{d}}
\newcommand{\dint}[1]{\mathchoice{\!\plaind#1\,}{\!\plaind#1\,}{\!\plaind#1\,}{\!\plaind#1\,}}

\newcommand{\pdf}[2]{\mathcal{P}^{\supmarker{#1}}\left(#2\right)}

\makeatletter
\newcommand{\supmarker}[1]{{\@ifempty{#1}{}{(#1)}}}
\makeatother

\newcommand{\elabel}[1]{\label{eq:#1}}
\newcommand{\eref}[1]{(\ref{eq:#1})}
\newcommand{\Eref}[1]{Eq.~(\ref{eq:#1})}

\newcommand{\slabel}[1]{\label{sec:#1}}

\newcommand{\Sref}[1]{Sec.~\ref{sec:#1}}
\newcommand{\Aref}[1]{Appendix~\ref{sec:#1}}

\newcommand{\flabel}[1]{\label{fig:#1}}
\newcommand{\fref}[1]{Fig.~\ref{fig:#1}}
\newcommand{\Fref}[1]{Fig.~\ref{fig:#1}}

\newcommand{\latin}[1]{{\it #1}}
\newcommand{\ie}{\latin{i.e.}\@\xspace}
\newcommand{\eg}{\latin{e.g.}\@\xspace}

\newcommand{\etal}{\latin{et~al}.\@\xspace}

\newcommand{\Exp}[1]{\operatorname{exp}\left(#1\right)}
\renewcommand{\exp}[1]{\mathchoice{e^{#1}}{\operatorname{exp}\left(#1\right)}{\operatorname{exp}\left(#1\right)}{\operatorname{exp}\left(#1\right)}}
\newenvironment{subeqnarray}[1]{\begin{subequations}#1\begin{eqnarray}}{\end{eqnarray}\end{subequations}\ignorespacesafterend}

\usepackage{subfigure}

\begin{document}

\title{The perils of thresholding}

\author{Francesc Font-Clos}
\email{fontclos@crm.cat}
\affiliation{Centre de Recerca Matem{\`a}tica, Edifici C, Campus
Bellaterra, E-08193 Bellaterra, Barcelona, Spain}
\affiliation{Department de Matem{\`a}tiques, Universitat Aut{\`o}noma de Barcelona,
Edifici C, E-08193 Bellaterra, Barcelona, Spain}

\author{Gunnar Pruessner}
\email{g.pruessner@imperial.ac.uk}
\affiliation{Department of Mathematics,
Imperial College London,
180 Queen's Gate,
London SW7 2BZ, United Kingdom}

\author{Anna Deluca}
\email{adeluca@pks.mpg.de}
\affiliation{Max Planck Institute for the
Physics of Complex Systems,
N{\"o}thnitzer Stra{\ss}e 38,
D-01187 Dresden,
Germany}

\author{Nicholas R. Moloney}
\email{n.moloney@lml.org.uk}
\affiliation{London Mathematical Laboratory, 14 Buckingham Street,
London WC2N 6DF, United Kingdom}

\begin{abstract}
The thresholding of time series of activity or intensity is frequently
used to define and differentiate events. This is either implicit, for
example due to resolution limits, or explicit, in order to filter
certain small scale physics from the supposed true asymptotic
events. Thresholding the birth-death process, however, introduces a
scaling region into the event size distribution, which is
characterised by an exponent that is unrelated to the actual asymptote
and is rather an artefact of thresholding. As a result, numerical fits
of simulation data produce a range of exponents, with the true
asymptote visible only in the tail of the distribution. This tail is
increasingly difficult to sample as the threshold is increased.  In
the present case, the exponents and the spurious nature of the scaling
region can be determined analytically, thus demonstrating the way in
which thresholding conceals the true asymptote. The analysis also
suggests a procedure for detecting the influence of the threshold by
means of a data collapse involving the threshold-imposed scale.
\end{abstract}

\maketitle

\section{Introduction}
Thresholding is a procedure applied to (experimental) data either
deliberately, or effectively because of device limitations. The
threshold may define the onset of an event and/or an effective zero,
such that below the threshold the signal is regarded as $0$. An
example of thresholding is shown in
\Fref{illustration_time_series_bd}.  Experimental data often comes
with a detection threshold that cannot be avoided, either because the
device is insensitive below a certain signal level, or because the
signal cannot be distinguished from noise. The quality of a
measurement process is often quantified by the noise to signal ratio,
with the implication that high levels of noise lead to poor
(resolution of the) data. Often, the rationale behind thresholding is
to weed out small events which are assumed irrelevant on large scales,
thereby retaining only the asymptotically big events which are
expected to reveal (possibly universal) large-scale physics.

\begin{figure}
\resizebox{0.49\textwidth}{!}{
\includegraphics{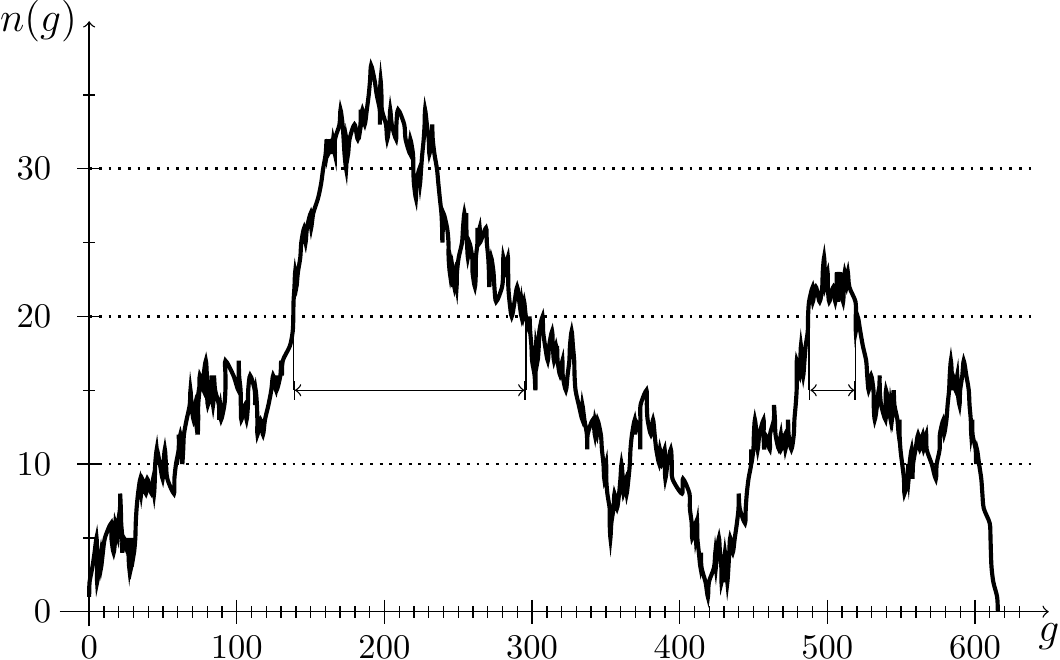}
}
\caption{\flabel{illustration_time_series_bd} Example of thresholding
  of a time series. An event begins when the signal exceeds the
  threshold (dotted lines, $h=10,20,30$) and ends as soon as the
  signal falls below the threshold. Increasing levels of the threshold
  lead (non-monotonically) to different numbers of events and,
  provided the signal eventually ends, monotonically smaller total
  event durations. The main focus of this paper is on the statistics
  of the individual event durations, as exemplified by the two
  intervals for the intermediate threshold.
}
\end{figure}

Most, if not all, of physics is due to some basic interactions that
occur on a ``microscopic length scale'', say the interaction between
water droplets or the van der Waals forces between individual water
molecules.  These length scales separate different
realms of physics, such as between micro-fluidics and molecular physics or
between molecular physics and atomic physics. However, these are
\emph{not} examples of the thresholds we are concerned with in the
following. Rather, we are interested in an often arbitrary microscopic
length scale well above the scale of the microscopic physics that
governs the phenomenon we are studying, such as the spatiotemporal
resolution of a radar observing precipitation (which is much coarser than
the scale set by microfluidics), or the resolution of the magnetometer
observing solar flares, (which is much coarser than the scale set by
atomic physics and plasma magnetohydrodynamics).

Such thresholds often come down to the device limitations of the
measuring apparatus, the storage facilities connected to it, or the
bandwidth available to transmit the data. For example, the earthquake
catalogue of Southern California is only complete above magnitude $3$,
even though the detection-threshold is around magnitude $2$
\cite{schorlemmer2008probability}. One fundamental problem is the
noise-to-signal ratio mentioned above. Even if devices were to improve
to the level where the effect of noise can be disregarded,
thresholding may still be an integral part of the measurement. For
example, the distinction between rainfall and individual drops
requires a separation of microscale and macroscale which can be highly
inhomogeneous \cite{lovejoy2003large}. Solar flares, meanwhile, are
defined to start when the solar activity exceeds the threshold and
end when it drops below, but the underlying solar activity never
actually ceases~\cite{Paczuski_btw}.

Thresholding has also played an important r{\^o}le in theoretical
models, such as the Bak-Sneppen Model \citep{BakSneppen:1993} of
Self-Organised Criticality \cite{Pruessner:2012:Book}, where the
scaling of the event-size distribution is a function of the threshold
\citep{PaczuskiMaslovBak:1996} whose precise value was the subject of
much debate \citep{Sneppen:1995b,Grassberger:1995}. Finite size
effects compete with the threshold-imposed scale, which has been used
in some models to exploit correlations and predict extreme events
\cite{Garber_pre}.

Often, thresholding is tacitly assumed to be ``harmless'' for the
(asymptotic) observables of interest and beneficial for the numerical
analysis. We will argue in the following that this assumption may be
unfounded: the very act of thresholding can distort the data and the
observables derived from it.  To demonstrate this, we will present an
example of the effect of thresholding by determining the apparent
scaling exponents of a simple stochastic process, the birth-death
process (BDP). We will show that thresholding obscures the asymptotic
scaling region by introducing an additional prior scaling region,
solely as an artefact. Owing to the simplicity of the process, we can
calculate the exponents, leading order amplitudes and the crossover
behaviour analytically, in excellent agreement with simulations. In
doing so, we highlight the importance of sample size since, for small
samples (such as might be accessible experimentally), only the
``spurious'' threshold-induced scaling region that governs the process
at small scales may be accessible. Finally, we discuss the
consequences of our findings for experimental data analysis, where
detailed knowledge of the underlying process may not be available,
usually the mechanism behind the process of interest is unclear, and
hence such a detailed analysis is not feasible. But by attempting a
data collapse onto a scaling ansatz that includes the
threshold-induced scale, we indicate how the effects of thresholding
can be revealed.

\begin{figure}
\subfigure[\hspace*{0.5mm}Small sample size $\NC=10^3$. \flabel{pdf_th_100_intro_small}]{\includegraphics[width=0.48\textwidth]{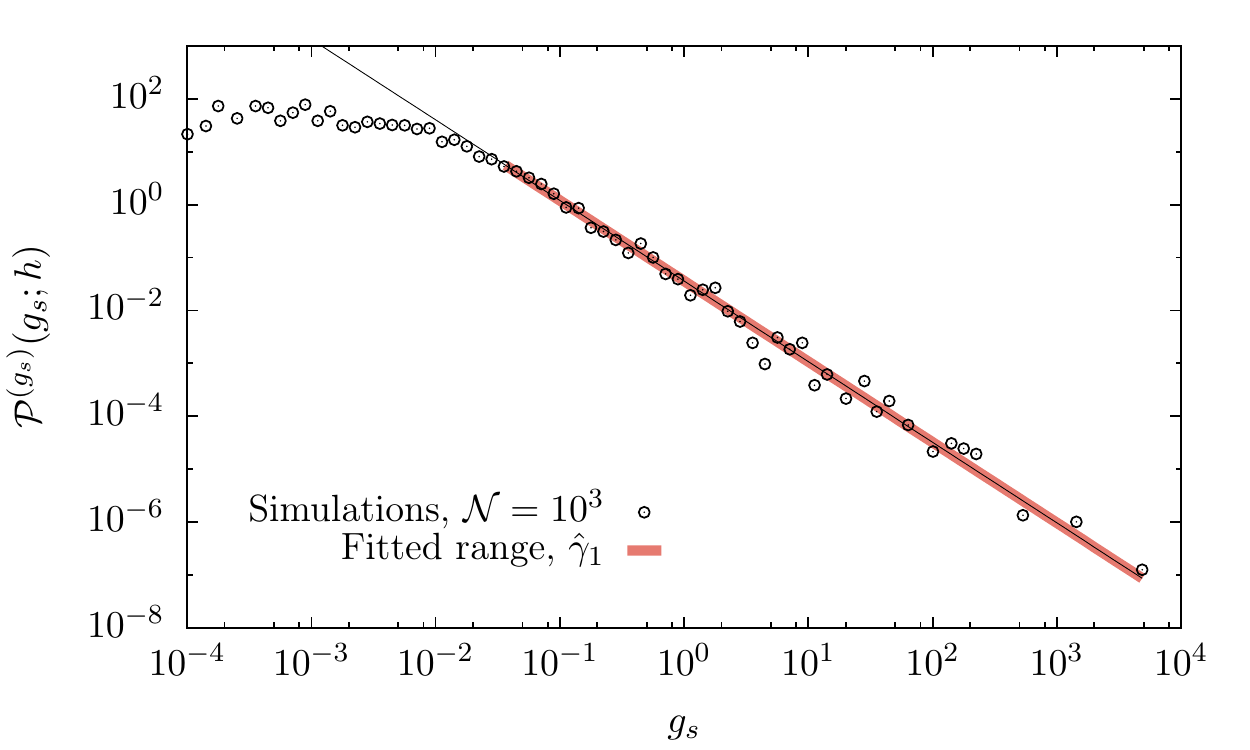}}
\subfigure[\hspace*{0.5mm}Large sample size $\NC=10^{10}$.  \flabel{pdf_th_100_intro_large}]{\includegraphics[width=0.48\textwidth]{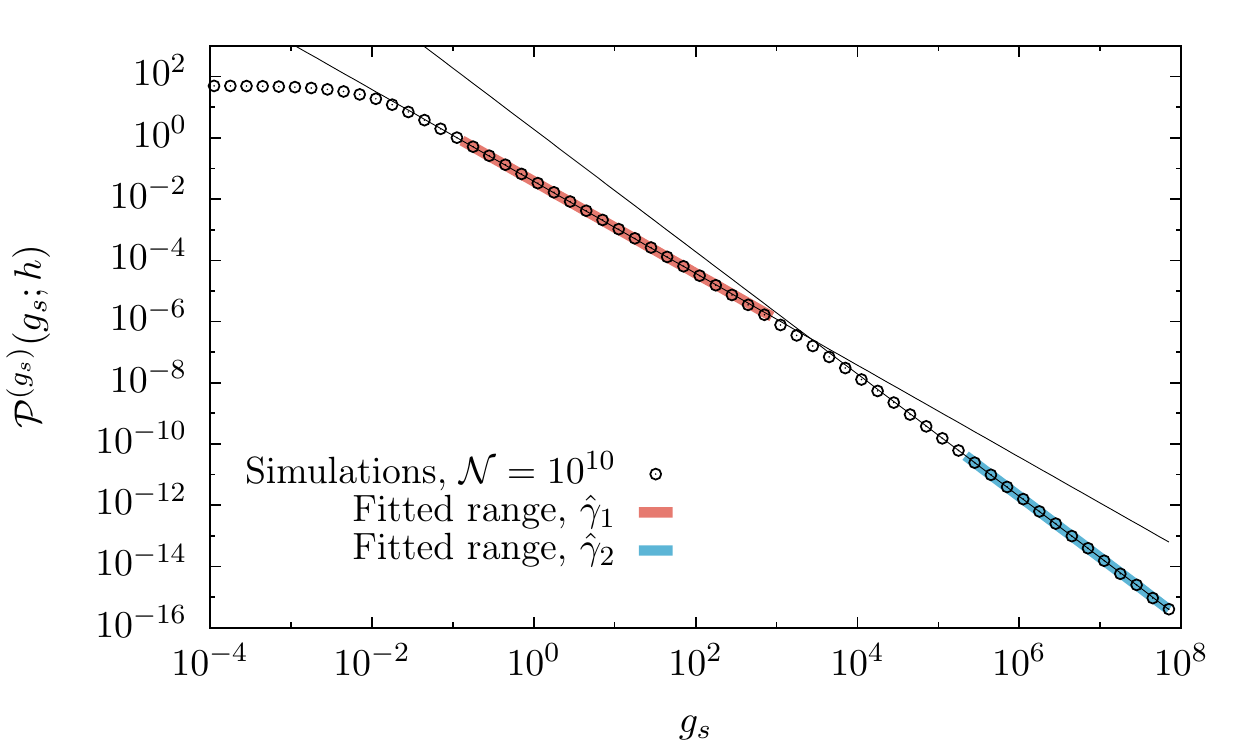}}
\caption{\flabel{pdf_th_100_intro}
\subref{fig:pdf_th_100_intro_small}: The PDF $\pdf{g_s}{g_s;h}$ of the
survival time $g_s$ of a thresholded BDP, with a threshold of $h=100$,
estimated from Monte Carlo simulations using a limited sample size of
$\NC=10^3$. Fitting a power law yields an
exponent of  $\hat\gamma_1=1.52(3)$ over the range $[0.031, 1.259\cdot 10^5]$ ,
with a $p$-value of 0.71. 
\subref{fig:pdf_th_100_intro_large}: Same as above, but using a sample
size of $\NC=10^{10}$.
In this case, two power laws can be fitted in two different regimes: below 
$g_\cross=8\pi h$, we find $\hat\gamma_1=1.50070(2)$ in the (fixed) range $[10^{-2},10^{3}]$, while above $g_\cross$, the fit leads to
 $\hat\gamma_2=1.998(4)$ over the range $[1.99\cdot 10^5,3.16 \cdot 10^8]$ , with a p-value of 0.55. Monte Carlo simulations are shown as symbols, while the small (large) regime
 power-law fit is plotted with full black lines, and the fitted range marked with red (blue) shading.}
\end{figure}

The outline of the paper is as follows: In \Sref{model} we introduce
the model and the thresholding applied to it. To illustrate the
problems that occur when thresholding real data, we analyse in detail
some numerical data. The artefact discovered in this analysis finds
explanation in the theory present in \Sref{results}. We discuss these
findings and suggest ways to detect the problem in the final section.

\section{Model}\slabel{model}

In order to quantify numerically and analytically the effect of
thresholding, we study the birth-death \cite{Gardiner:1997} process
(BDP) with Poissonian reproduction and extinction rates that are
proportional to the population size. More concretely, we consider the
population size $n(g)$ at (generational) time $g\ge0$. Each individual
in the population reproduces and dies with the same rate of $1/2$ (in
total unity, so that there are $n(g)$ birth or death events or
``updates'' per time unit on average); in the former case (birth) the
population size increases by $1$, in the latter (death) it decreases
by $1$. The state $n(g)=0$ is absorbing
\cite{Hinrichsen:2000a}. Because the instantaneous rate with which the
population $n(g)$ evolves is $n(g)$ itself, the exponential
distributions from which the random waiting times between events are
drawn are themselves parameterised by a random variable, $n(g)$.

Because birth and death rates balance each other, the process is said
to be at its critical point \cite{Harris:1963}, which has the peculiar
feature that the expectation of the population is constant in time,
$\ave{n(g)}=n(g_0)$, where $\ave{\cdot}$ denotes the expectation and
$n(g_0)$ is the initial condition, set to unity in the following. This
constant expectation is maintained by increasingly fewer surviving
realisations, as each realisation of the process terminates almost
surely. We therefore define the survival time as the time $g_s-g_0$
such that $n(g)>0$ for all $g_0 \le g<g_s$ and $n(g)=0$ for all $g\ge
g_s$.  For simplicity, we may shift times to $g_0=0$, so that $g_s$
itself is the survival time. It is a continuous random variable, whose
probability density function (PDF) is well known to have a power law
tail in large times, $\pdf{g_s}{g_s}\propto g_s^{-2}$ \citep[][as in
  the branching process]{Harris:1963}.

In the following, we will introduce a threshold, which mimics the
suppression of some measurements either intentionally or because of
device limitations. For the BDP this means that the population size
(or, say, ``activity'') below a certain, prescribed level, $h$, is
treated as $0$ when determining survival times.  In the spirit of
\citep[][also solar flares,
  \citealp{Paczuski_btw}]{PetersHertleinChristensen:2002}, the
threshold allows us to distinguish events, which, loosely speaking,
start and end whenever $n(g)$ passes through $h$.

Explicitly, events start at $g_0$ when
$\lim_{\epsilon\to0^+}n(g_0-\epsilon)=h$ and $n(g_0)=h+1$. They end at
$g_s$ when $n(g_s)=h$, with the condition $n(g)>h$ for all $g_0\le
g<g_s$.  This is illustrated in
Figs.~\ref{fig:illustration_time_series_bd} and
\ref{fig:illustration_time_series_bd_zoom}. No thresholding takes
place (\ie the usual BD process is recovered) for $h=0$, in which case
the initial condition is $n(g_0)=1$ and termination takes place at
$g_s$ when $n(g_s)=0$. For $h>0$ one may think of $n(g)$ as an
``ongoing'' time series which never ceases and which may occasionally
``cross'' $h$ from below (starting the clock), returning to $h$ some
time later (stopping the clock). In a numerical simulation one would
start $n(g)$ from $n(g_0)=h+1$ at $g_0=0$ and wait for $n(g)$ to
arrive at $n(g)=h$ from above. The algorithm may be summarised as
\begin{algorithmic}
\For{$i=1\dots \NC$}
	\State $n\gets h+1$
	\State $g_i\gets 0$
	\While{$n > h$}
		\State $g_i \gets g_i +\xi(n)$
		\State $n\gets n+b$
	\EndWhile
\EndFor
\end{algorithmic}
where $\xi(n)$ is an exponential random variable with rate $n$, and $b$
stands for a random variable that takes the values $\{-1,1\}$ with
probability 1/2. In our implementation of the algorithm, all random
variables are handled with the \textsf{GNU Scientific Library}
\citep{GalassiETAL:2009}. 

\subsection{Numerics and data analysis}
Monte-Carlo runs of the model reveal something unexpected: The
exponent of the PDF of the thresholded BDP appears to change from
$\pdf{g_s}{g_s}\propto g_s^{-2}$ at $h=0$ to $\pdf{g_s}{g_s}\propto
g_s^{-3/2}$ at $h=100$ or, in fact, any reasonably large
$h\gtrsim10$. \Fref{pdf_th_100_intro} shows $\pdf{g_s}{g_s}$ for the
case of $h=100$ and two different sample sizes, $\NC_1=10^{3}$ and
$\NC_2=10^{10}$, corresponding to ``scarce data'' and ``abundant
data'', respectively. In the former case, the exponent of the PDF is
estimated to be $\hat\gamma_1=1.52(3) \approx 3/2$; in the latter, the
PDF splits into two scaling regimes, with exponents
$\hat\gamma_1=1.50070(2)\approx 3/2$ and
$\hat\gamma_2=1.998(4)\approx 2$.  This phenomenon can be investigated
systematically for different sample sizes $\NC$ and thresholds $h$.

We use the fitting procedure introduced in \citet{DelucaCorral}, which
is designed not only to estimate the exponent, but to determine the
range in which a power law holds in an objective way. It is based on
maximum likelihood methods, the Kolmogorov-Smirnov test and Monte
Carlo simulations of the distributions.  In \fref{N_gamma2} we show
the evolution of the estimated large scale exponent, $\hat\gamma_2$,
for different $\NC$ and for different $h$. The fits are made by
assuming that there is a true power law in a finite range [a,b].
For values of
the exponent between 1.5 and 2 larger error bars are observed. For
these cases, less data is fitted but the fitting range is always at
least two orders of magnitude wide.
 
It is clear from \fref{N_gamma2} that $\NC$ has to be very large in order
to see the true limiting exponent.  Even the smallest $h$ investigated,
$h=20$, needs a sample size of at least $\NC=10^7$, while for $h=5\,000$
the correct exponent is not found with less than about $\NC=10^{10}$.

The mere introduction of a threshold therefore changes the PDF of
events sizes significantly. It introduces a new, large scaling regime,
with an exponent that is misleadingly different from that
characterising large scale asymptotics. In fact, for small sample
sizes ($\NC_1=10^3$, see \fref{pdf_th_100_intro_small}), the only
visible regime is that induced by thresholding (in our example,
$\gamma_1=3/2$), while the second exponent ($\gamma_2=2$), which, as
will be demonstrated below, governs the large scale asymptotics,
remains hidden unless much larger sample sizes are used
(\Fref{pdf_th_100_intro_large}).

Although the algorithm is easy to implement, finding the two scaling
regimes numerically can be challenging. There are a number of
caveats:
\begin{enumerate}
\item[(1)] The crossover point $g_\cross$ between the two scaling
  regimes scales linearly with the threshold, $g_\cross = 8\pi h$ (see
  \Sref{distribution_of_gs}), effectively shifting the whole
  $g_s^{-2}$ asymptotic regime to larger and thus less likely values
  of $g_s$. To maintain the same \emph{number} of events above
  $g_\cross \propto h$, one needs $\NC\int_{g_\cross}^{\infty}
  \dint{g_s} g_s^{-2}=\text{const}$, \ie $\NC\propto h$.
\item[(2)] Because the expected running time of the algorithm
  diverges, one has to set an upper cutoff on the maximum generational
  timescale, say $g_s< G$.  If the computational complexity for each
  update is constant, an individual realisation, starting from
  $n(0)=h+1$ and running up to $n(g_s)=h$ with $g_s<G$, has complexity
  $\OC(g_s^2)$ in large $g_s$ where $g_s^2$ is the scaling of the
  expected survival time of the mapped random walker introduced
  below. The expected complexity of realisations that terminate before
  $G$ (with rate $\sim1/g_s^2$) is therefore linear in $G$,
  $\int_1^{G}\dint{g_s} g_s^{-2} g_s^2 = G-1$.  With the random walker
  mapping it is easy to see that the expected population size $n(g)$
  of realisations that terminate after $G$ (and therefore have to be
  discarded as $g_s$ exceeds $G$) is of the order $n(g_s)\sim G$ for
  $g_s=G$. These realisations, which appear with frequency $\propto
  1/G$, have complexity $\OC(G^2)$, \ie the complexity of realisations
  of the birth-death process is $\OC(G)$ both for those counted into
  the final tally and those dismissed because they exceed $G$. There is no
  point probing beyond $G$ if $\NC$ is too small to produce a
  reasonable large sample on a logarithmic scale, $\NC
  \int_G^{2G}\dint{g_s} g_s^{-2}=\text{const}$, so that $\NC \sim G$
  and thus the overall complexity of a sample of size $\NC$ is
  $\OC(\NC^2)$ and thus $\OC(h^2)$ for $G\sim g_\cross \sim h$ and
  $\NC\propto h$ from above.
\end{enumerate}

That is, larger $h$ necessitates larger $\NC$, leading to
\emph{quadratically} longer CPU time. In addition, parallelisation of
the algorithm helps only up to a point, as the (few) biggest events
require as much CPU time as all the smaller events taken together. The
combination of all these factors has the unfortunate consequence that,
for large enough values of $h$, observing the $\pdf{g_s}{g_s} \propto
g_s^{-2}$ regime is simply out of reach (even for moderate values of
$h$, such as $h=100$, to show the crossover as clearly as in
\fref{pdf_th_100_intro}, a sample size as large as $\NC=9\cdot 10^{9}$ was
necessary, which required about 1810 hours of CPU time).

\begin{figure} 
\includegraphics[width=0.48\textwidth, angle=0]{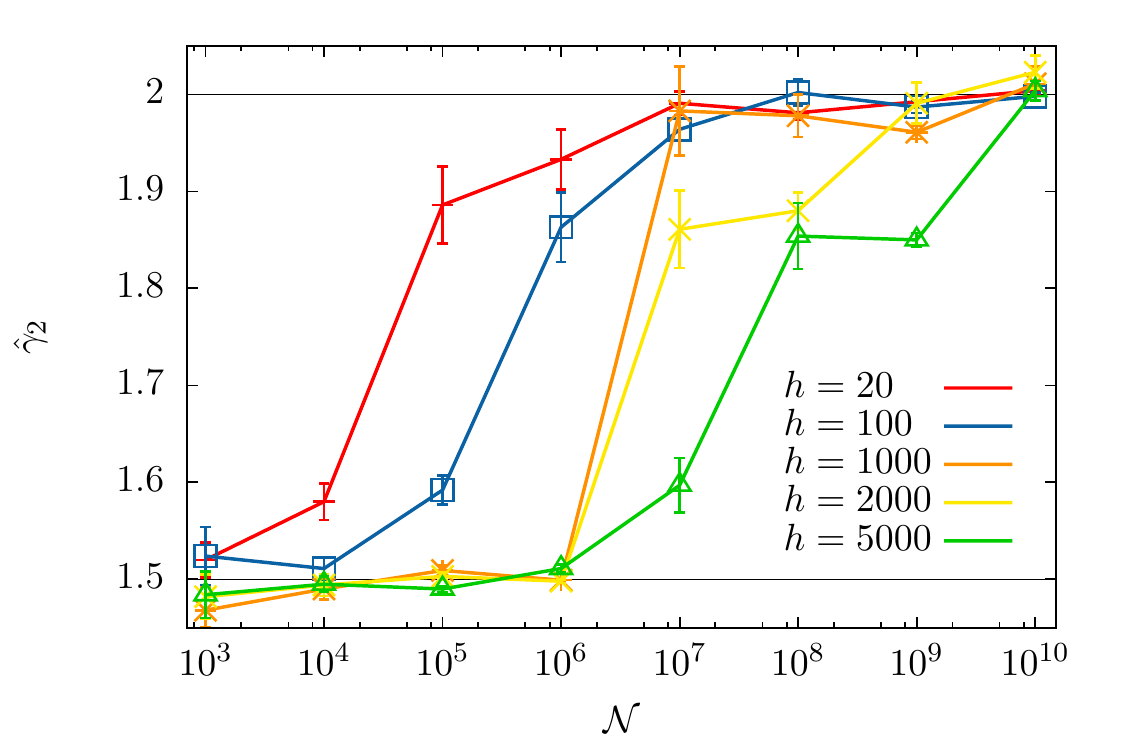} 
\caption{Estimated large scale exponent $\hat\gamma_2$ for different
thresholds $h$ and sample size $\NC$. The error bars correspond to
one standard deviation and are inversely proportional to the number of
data within the fitted range. \flabel{N_gamma2}} 
\end{figure}

\section{Results}\slabel{results}
While it is straightforward to set up a recurrence relation for the
generating function if the threshold is $h=0$, the same is not true
for $h>0$. This is because the former setup ($h=0$) does not require
an explicit implementation of the absorbing wall since the process
terminates naturally when $n(g)=0$  (there
is no individual left that can reproduce or die). If, however, $h>0$,
the absorbing wall has to be treated explicitly and that is difficult
when the evolution of the process (the effective diffusion constant)
is a function of its state, \ie the noise is multiplicative.  In
particular, a mirror charge trick cannot be applied.

\begin{figure}
\resizebox{0.45\textwidth}{!}{
\includegraphics{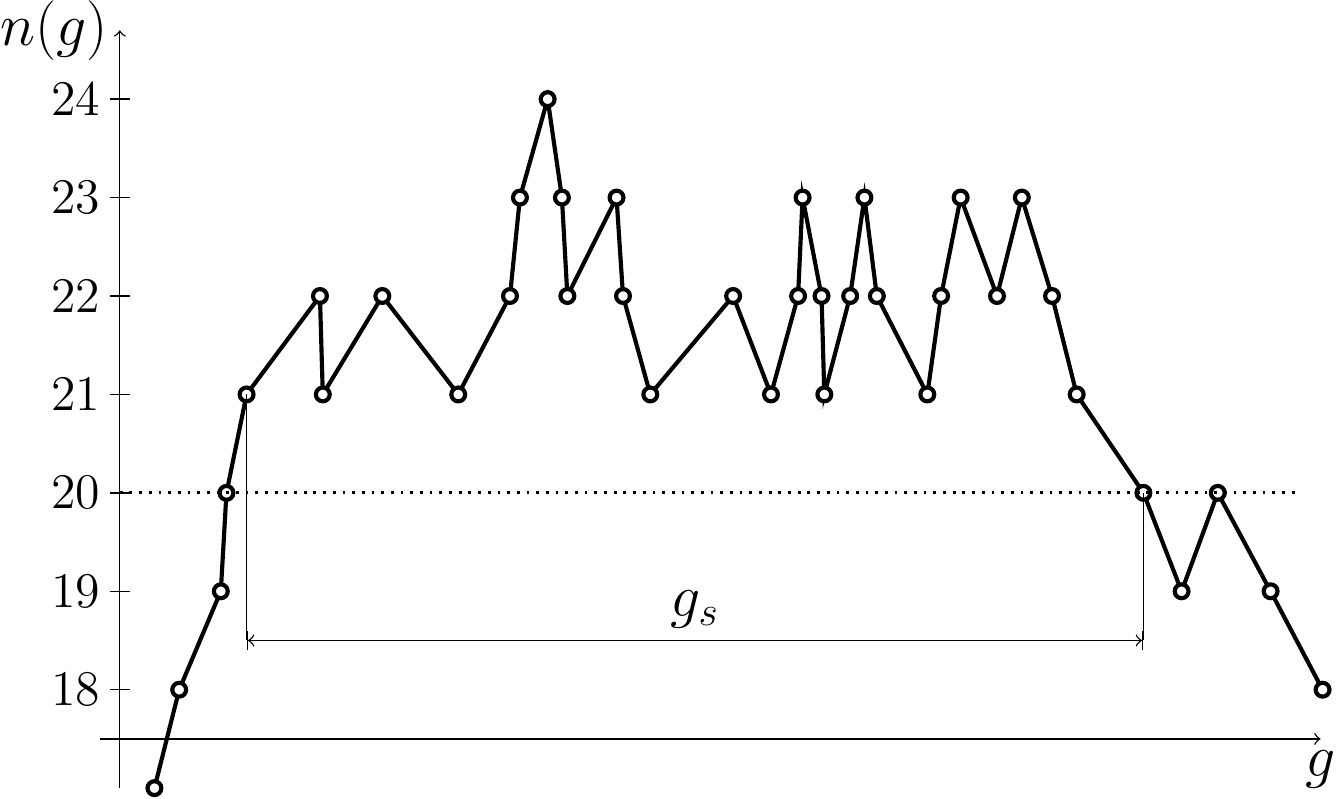}
}
\caption{\flabel{illustration_time_series_bd_zoom}
Magnification of the right interval in 
\Fref{illustration_time_series_bd}. The clock starts when $n(g)$ exceeds
the threshold and stops when $n(g)$ returns to the threshold.}
\end{figure}
\begin{figure}
\resizebox{0.45\textwidth}{!}{
\includegraphics{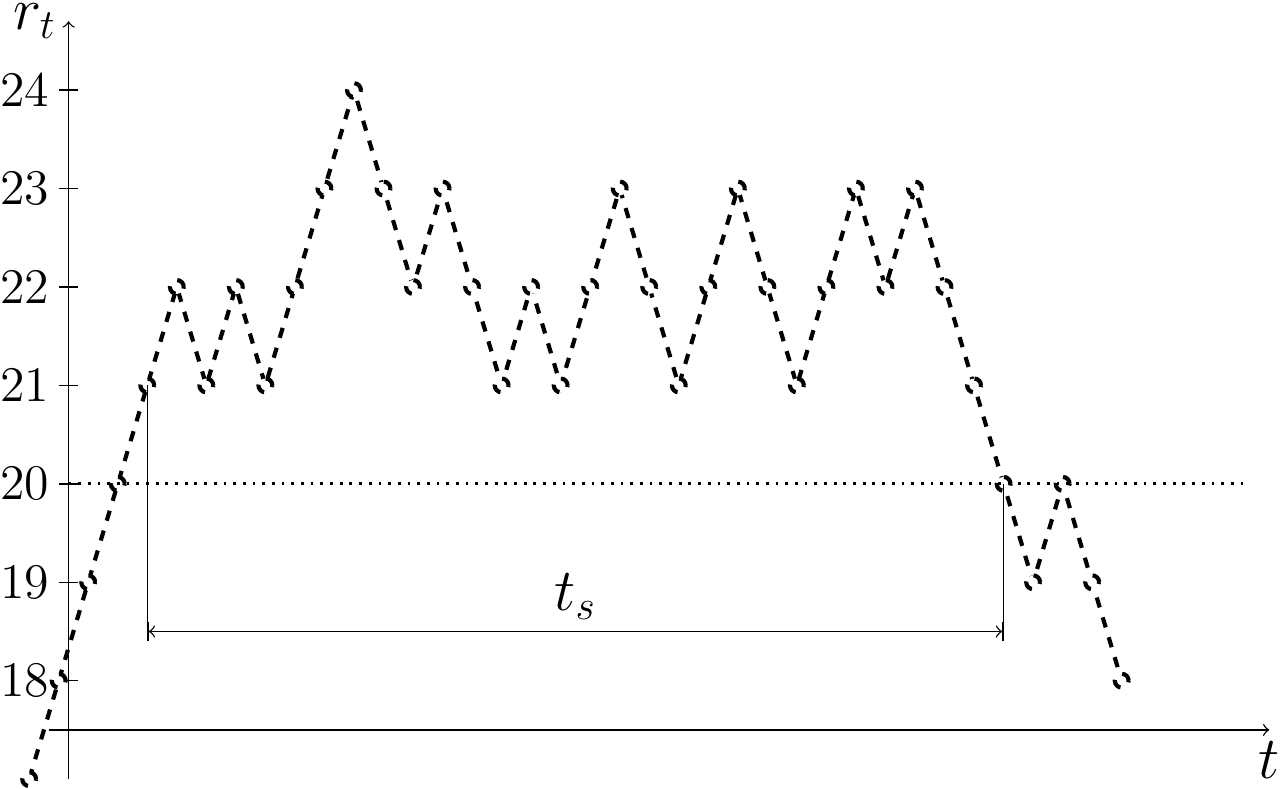}
}
\caption{\flabel{illustration_time_series_bd_zoomMAP}
The same data as in \Fref{illustration_time_series_bd_zoom} but on the
mapped time scale of the random walker, which evolves in equally spaced,
discrete steps. The survival time is necessarily odd, $t_s=2T-1$, $T\in
\Nset$ ($t_s=29$ in this example).
}
\end{figure}

However, the process can be mapped to a simple random walk by ``a
change of clocks'', a method detailed in
\cite{RubinPruessnerPavliotis:2014}.  For the present model, we
observe that $n(g)$ performs a fair random walk $r_t$ by a suitable
mapping of the generational timescale $g$ to that of the random
walker, $r_t(g)=n(g)$ with $t(g)\in\Nset$. In fact, because of the
Poissonian nature of the BD process, birth and death almost surely
never occur simultaneously and a suitable, unique $t(g)$ is found by
$t(0)=0$ and
\begin{equation}
\lim_{\epsilon\to0^+} t(g+\epsilon)-t(g-\epsilon) = 
\lim_{\epsilon\to0^+} | n(g+\epsilon)-n(g-\epsilon)|
\end{equation}
\ie $t(g)$ increases whenever $n(g)$ changes and is therefore an
increasing function of $g$. With this map, $r_t$ is a simple random
walk along an absorbing wall at $h$, see
\Fref{illustration_time_series_bd_zoomMAP}. The challenge is to derive
the statistics of the survival times $g_s$ on the time scale of the BD
process from the survival times $t_s$ on the time scale of the random
walk.

In the following, we first approximate some important properties of the
survival times in a handwaving manner before presenting a mathematically
sound derivation in \Sref{maths}.

\subsection{Approximation}
\slabel{approximation} 
The expected waiting time\footnote{In a numerical simulation this
  would be the time increment.} between two events in the BDP is
$1/n$, if $n$ is the current population size, with $n=n_x+h$ such that
$n_x$ is the excess of $n$ above $h$. As discussed in detail in
\Sref{maths}, $n_x$ is a time-dependent random variable, and so taking
the ensemble average of the waiting time is a difficult task.  But on
the more convenient time scale $t$, the excess $n_x$ performs a random
walk and it is in that ensemble, with that time scale, where we
attempt to find the expectation
\begin{equation}
\spave{g_s(t_s;h)}=\sum_{t=0}^{t_s-1}
\ave{\frac{1}{n_x(t)+h}}_{\RC(t_s)} \ ,
\elabel{gs_of_ts}
\end{equation}
which is the expected survival time of a thresholded BD process given
a certain return (or survival) time $t_s$ of the random walker. In
this expression $n_x(t)$ is a time-dependent random variable and the
ensemble average $\ave{\cdot}_{\RC(t_s)}$ is taken over all random
walker trajectories $\RC(t_s)$ with return time $t_s$.  To ease
notation, we will include the argument of $\RC(t_s)$ only where
necessary. Approximating the random variable $g_s$ by its mean
$\spave{g_s(t_s;h)}$ given in \Eref{gs_of_ts} and approximated further
below affords an approximate map of the known PDF $\pdf{t_s}{t_s}$ of
$t_s$ to the PDF $\pdf{g_s}{g_s}$ of $g_s$,
\begin{equation}
\pdf{g_s}{g_s} \frac{\plaind}{\plaind t_s} \spave{g_s(t_s;h)} 
\approx
\pdf{t_s}{t_s}
\end{equation}
This map will be made rigorous in \Sref{maths}, avoiding the use of
$\spave{g_s(t_s;h)}$ in lieu of the random variable.

In a more brutal approach, one may approximate the time dependent
excess $n_x(t)$ in \Eref{gs_of_ts} by its expectation conditional to a
certain survival time $t_s$,
\begin{multline}
\ave{\frac{1}{h+n_x(t)}}_{\RC} 
= \frac{1}{h+\ave{n_x(t)}_{\RC}}
\ave{\frac{1}{1+\frac{n_x(t)-\ave{n_x(t)}_{\RC}}{h+\ave{n_x(t)}_{\RC}}}}\\
= \frac{1}{h+\ave{n_x(t)}_{\RC}} + \text{(higher order terms)}
\elabel{dropping_hot}
\end{multline}
so that the expected survival time $g_s(t_s)$ given a certain return
time $t_s$ is approximately $t_s/(h+\ave{n_x(t)}_{\RC})$.

The quantity $\ave{n_x(t)}_{\RC}$ is the expected excursion of a
random walker, which is well-known to be
\begin{equation}
\ave{n_x(t)}_{\RC}\approx\sqrt{\frac{\pi}{8}}t_s^{1/2}
\end{equation}
in the continuum limit (with diffusion constant $1/2$)
\citep[\eg][]{MajumdarComtet:2004,MajumdarComtet:2005}. Thus,
\begin{equation}
\spave{g_s(t_s;h)}\approx \frac{t_s}{h+\sqrt{\pi t_s/8}} \ .
\elabel{approx_gs}
\end{equation}
At small times, $h\gg\sqrt{\pi t_s/8}$, the relation between $g_s$ and
$t_s$ is essentially linear, $g_s\approx t_s/h$, whereas for large
times, $h\ll\sqrt{\pi t_s/8}$, the asymptote is $g_s\approx \sqrt{8
  t_s/\pi}$. Writing the right-hand-side of \Eref{approx_gs} in the
form $\sqrt{8 t_s/\pi} \frac{1}{1+\sqrt{8 h^2/(\pi t_s)}}$ allows us
to extract the scaling of the crossover time. The argument of the
square root is of order unity when $t_{\cross}=8 h^2/\pi$, for which
$g_s(t_{\cross},h) \approx 4h/\pi$. Moreover, one can read off the
scaling form
\begin{equation}
\spave{g_s(t_s;h)} \approx t_s^{1/2} \GC(t_s/h^2) \ ,
\elabel{approx_gs_scaling}
\end{equation}
with $\GC(x)=\sqrt{8/\pi}/(1+\sqrt{8/(\pi x)})$ and asymptotes
$\GC(x)\approx \sqrt{x}$ for small $x$ and
$\lim_{x\to\infty}\GC(x)=\sqrt{8/\pi}$.

The PDF of the survival time 
\begin{equation}
\pdf{t_s}{t_s} = \frac{1}{\sqrt{4 \pi D t}} \frac{a}{D t_s} \exp{-\frac{a^2}{4 D t_s}}
\end{equation}
of a random walker along an absorbing wall is well-known to be a power
law $\propto t_s^{-3/2}$ for times $t_s$ large compared to the time
scale set by the initial condition, \ie the distance $a$ of the random
walker from the absorbing wall at time $t=0$. The precise value of $a$
is effectively determined by the details the continuum approximation,
here $a=1$, $D=1/2$, and so we require $1\ll 2t_s$.

To derive the PDF of the BD process, note that \Eref{approx_gs} has
the unique inverse
$t_s(g_s)=\frac{\pi g_s^2}{16} \TC(\frac{16 h}{\pi g_s})$,
where $\TC(y)=1+y+\sqrt{1+2y}$. Evaluating the crossover time by
setting $y=1$ yields $g_{\cross}=16h/\pi$. The
PDF of the survival time of the BD process finally reads 
\begin{equation}
\pdf{g_s}{g_s;h} \sim 
\left(\frac{\pi}{16} \TC(y) \right)^{-1/2} g_s^{-2}
\left(2 - \frac{y \TC'(y)}{\TC(y)} \right)
\elabel{handwaving_BD_PDF}
\end{equation}
where $y=\frac{16 h}{\pi g_s}$. For small $y$, the last bracket
converges to $2$, so $\pdf{g_s}{g_s;h} \sim 2 \sqrt{8/\pi} g_s^{-2}$
for large $g_s$. For large $y$, the last bracket converges to $1$, so
$\pdf{g_s}{g_s;h} \sim (1/\sqrt{h}) g_s^{-3/2}$ for small $g_s$.

This procedure recovers the results in \Sref{maths}: For $g_s\ll
16h/\pi$ the PDF of the survival times in the BD process goes like
$g_s^{-3/2}$, and for $g_s\gg 16h/\pi$ like $g_s^{-2}$, independent of
$h$. \Eref{handwaving_BD_PDF} also gives a prescription for a
collapse, since $\pdf{g_s}{g_s;h} g_s^2$ plotted versus $g_s/h$
should, for sufficiently large $g_s$, reproduce the same curve, as
confirmed in \Fref{collapse} and \Fref{3regimes}.

Applying a threshold introduces a new scale, $16 h/\pi$, below which
the PDF displays a clearly discernible power law, $g_s^{-3/2}$,
corresponding to the return time of a random walker. The ``true''
$g_s^{-2}$ power law behaviour (the large $g_s$ asymptote) is visible
only well above the threshold-induced crossover.

\subsection{Detailed Analysis}\slabel{maths}
In the previous section we made a number of assumptions, in particular
the approximation of replacing the random variable by its expectation,
and the approximation in \Eref{dropping_hot}, which both require further
justification.

In the present section we proceed more systematically.  In particular,
we will be concerned with the statistics of the BD survival time
$g_s(\RC)$ \emph{given} a particular \emph{trajectory}
$\RC=\{r_0,r_1,\ldots,r_{t_s}\}$ of the random walk, where $t_s=2T-1$,
necessarily odd, $T\in\Nset$, see Figs.
\ref{fig:illustration_time_series_bd_zoomMAP} and
\ref{fig:rho_illustration}. We will then relax the constraint of the
trajectory and study the whole ensemble $\Omega$ of random walks
terminating at a particular time $2T-1$, denoting as $g_s(\Omega(T))$
a survival time drawn from the distribution of all survival times of a
BD process with a mapping to a random walker that terminates at $2T-1$
or, for simplicity, just $g_s(\Omega)$.  This will allow us to
determine the existence of a limiting distribution for
$g_s(\Omega)/\sqrt{T}$ and to make a quantitative statement about its
mean and variance. We will \emph{not} make any assumptions about the
details of that limiting distribution; in order to determine the
asymptotes of $\pdf{g_s}{g_s;h}$ we need only know that the limit
exists.

For a given trajectory $\RC$ of the random walk, the resulting
generational survival time $g_s(\RC)$ may be written as
\begin{equation}
\elabel{g_traj}
g_s(\RC) = \sum_{t=0}^{2T-2} \xi_t(r_t+h),
\end{equation}
where $\xi_t(\alpha)$ is a random variable drawn at time $t$ from an
exponential distribution with rate $\alpha$, \ie drawn from
$\alpha\exp{-\alpha\xi}$, and $r_t$ is the position of the random walk at time $t$, with initial
condition $r_0=1$ and terminating at $2T-1$ with $r_{2T-1}=0$ (see
\Fref{rho_illustration}). 

The mean and standard deviation of $\xi_t$ are $1/(r_t+h)$,
necessarily finite, so that by the central limit theorem the limiting
distribution of $g_s(\RC)/\sqrt{T}$ \emph{given} a trajectory $\RC$ is
Gaussian (for $T\gg 1$). This ensures that $g_s(\Omega)/\sqrt{T}$
has a limiting distribution (see \Aref{limiting_dist}).

It is straightforward to calculate the mean and standard deviation of
$g_s(\RC)$ for a particular trajectory $\RC$ that terminates after
$2T-1$ steps. Slightly more challenging is the mean $\mu(\Omega)$ and
variance $\sigma^2(\Omega)$ of $g_s(\Omega)$ for the entire ensemble
$\Omega$ of such trajectories.  The details of this calculation are
relegated to \Aref{mean_var_appendix}. Here, we state only the key
results. For the mean of the survival time, we find
\begin{equation}
\mu(\Omega)
\simeq
2 \sqrt{\pi T} +
2 h
\psi\left( \frac{h}{\sqrt{T}} \right)
\elabel{mu_Omega_largeT_mainText}
\end{equation}
(see \Eref{mu_Omega_largeT})
with $\psi(x)=\exp{-x^2} (\operatorname{Ei}(x)-\pi\EC(\imag x)/\imag)$ 
and asymptotes
\begin{subnumcases}{\mu(\Omega)\simeq\elabel{mu_asympt_mainText}}
2\sqrt{\pi T} & for $T\gg h^2$ \elabel{mu_asympt_mainText_Tggh}\\
2 T/h & for $T\ll h^2$ \elabel{mu_asympt_mainText_Tllh}
\end{subnumcases}
see \Eref{mu_asympt}. The variance is
\begin{equation}
\sigma^2(\Omega) \simeq  T\ \IC(x) - \mu(\Omega)^2 + \KC(x)
\elabel{sigma2_in_IKmu_mainText}
\end{equation}
(see \Eref{sigma2_in_IKmu})
with integrals $\IC(x)$ and $\KC(x)$ defined in \Eref{KI_integrals}
and with asymptotes
\begin{subnumcases}{\sigma^2(\Omega)\simeq\elabel{sigma2_asympt_mainText}}
4 \pi T \frac{\pi-3}{3} & for $T\gg h^2$ \\
2T/h^2 & for $T\ll h^2$ \ ,
\end{subnumcases}
see \Eref{sigma2_asympt}.  All these results are derived in the limit
$T\gg1$ in which the mapped random walker takes more than just a few
steps, corresponding to a continuum approximation. However, as shown
in the following, the results remain valid even for $T$ close to one.

To assess the quality of the continuum approximation and the validity
of the asymptotes, we extracted the mean $\mu(\Omega(T))$ and variance
$\sigma^2(\Omega(T))$ of the survival time $g_s(\Omega(T))$ from
simulated BDPs starting with a population size $n(0)=h+1$ and
returning to $n(g_s)=h$ after $2T-1$ updates (births or deaths), \ie
the process was conditioned to a particular value of $T$. In
particular, we set the threshold at $h=100$, and simulated a sample of
$10^5$ constrained BDPs for values $T=2^k, k=0\dots20$.  The results
are shown in \Fref{mu_sigma} and confirm the validity of the large
$T\gg1$ approximation in \Eref{mu_Omega_largeT_mainText} and
\Eref{sigma2_in_IKmu_mainText}, as well as the asymptotes
\Eref{mu_asympt_mainText} and
\Eref{sigma2_asympt_mainText}. Remarkably, as previously stated,
\Eref{mu_Omega_largeT_mainText} and \Eref{sigma2_in_IKmu_mainText} are
seen to be valid even when the condition $T\gg 1$ does not reasonably
hold.

\begin{figure}
\includegraphics[width=0.48\textwidth]{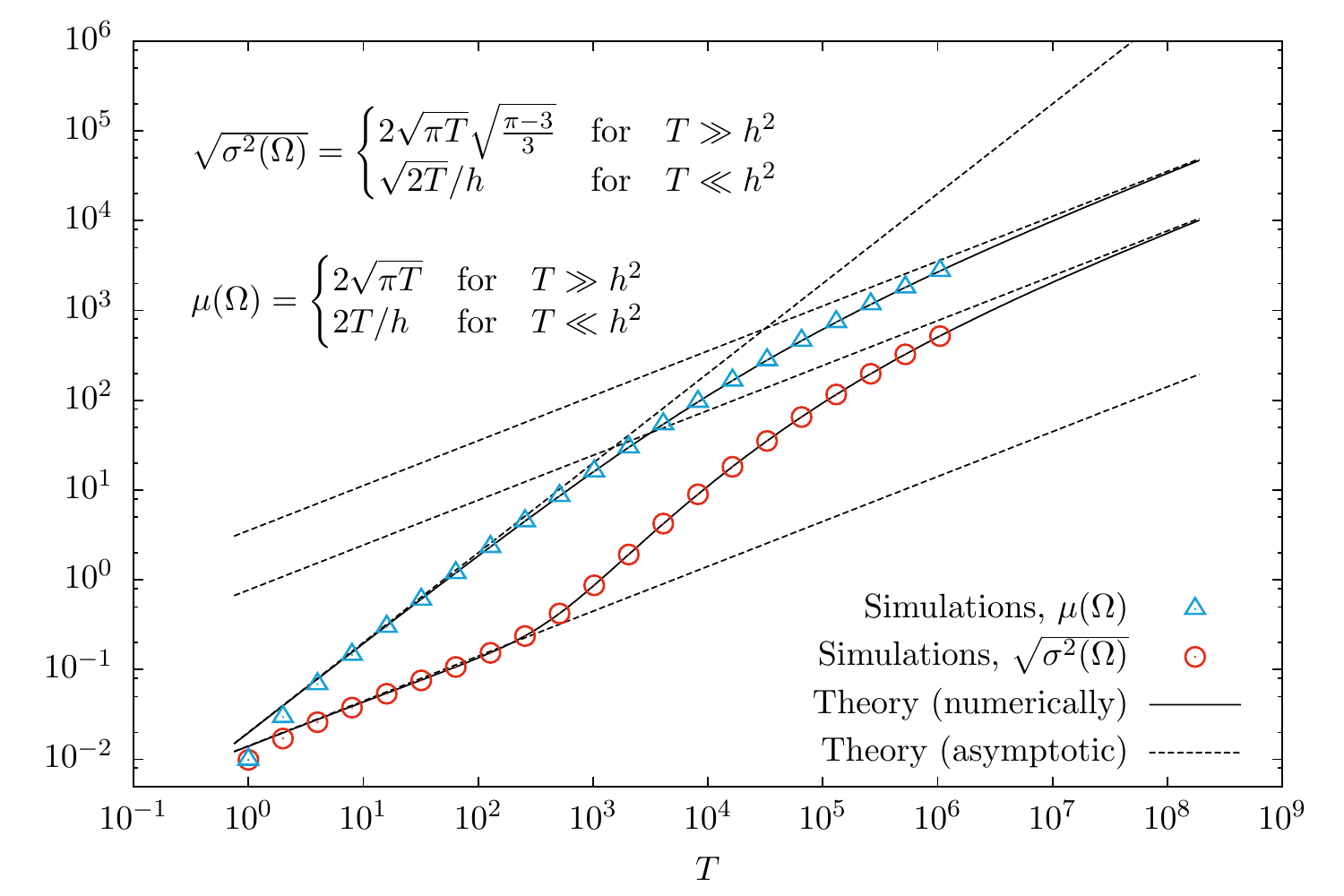}
\caption{\flabel{mu_sigma}
Numerical comparison of the approximations
\Eref{mu_Omega_largeT_mainText} and \Eref{sigma2_in_IKmu_mainText} (shown as
full lines), their asymptotes \Eref{mu_asympt_mainText} and
\Eref{sigma2_asympt_mainText}
(dashed) and the numerical estimates based on a sample of $10^5$
realisations per datapoint in a Monte-Carlo simulation
of a birth-death process  
constrained to $2T-1$ updates, with $h=100$ and $T=2^k, k=0\dots20$. 
}
\end{figure}

\subsubsection{Distribution of $g_s$}\slabel{distribution_of_gs}
For large $T$, the generational survival time $g_s$ \emph{given} a
survival time $2T-1$ of the mapped random walk has PDF
\begin{equation}\elabel{def_limiting_distribution}
\pdf{g_s}{g_s;h;T}\simeq
\frac{1}{\sqrt{\sigma^2(\Omega(T))}}\Phi\left(\frac{g_s-\mu(\Omega(T))}{\sqrt{\sigma^2(\Omega(T))}}\right),
\end{equation}
where $\Phi(x)$ denotes the limiting distribution of the rescaled
survival time $(g_s-\mu(\Omega(T)))/\sqrt{\sigma^2(\Omega(T)})$, and
the mean $\mu(\Omega(T))$ and variance $\sigma^2(\Omega(T))$ are given
by \Eref{mu_Omega_largeT_mainText} and
\Eref{sigma2_in_IKmu_mainText}. We demonstrate that $\Phi$ exists
and find its precise (non-Gaussian) form in \Aref{limiting_dist} for
completeness, but we will not use this result in what follows: to
extract the asymptotic exponents and first order amplitudes, see
below, knowledge of the mean $\mu(\Omega)$ and variance
$\sigma^2(\Omega)$ is sufficient.

As the ensembles $\Omega(T)$ are disjoint for different $T$, the
overall distribution $\pdf{g_s}{g_s;h}$ of survival generational times
is therefore given by the sum of the constrained distribution
$\pdf{g_s}{g_s;h;T}$ weighted by the probability of the mapped random
walk to terminate after $2T-1$ steps. In the limit of large $T$, as
assumed throughout, that weight is $ T^{-3/2}/(2\sqrt{\pi})$
\citep{Mohanty:1979}. Therefore,
\begin{equation}\elabel{p_of_g}
\pdf{g_s}{g_s;h}=\sum_{T=1}^{\infty}
\frac{T^{-3/2}}{2\sqrt{\pi}} \frac{1}{\sqrt{\sigma^2(\Omega(T))}} \Phi\left(\frac{g_s-\mu(\Omega(T))}{\sqrt{\sigma^2(\Omega(T))}}\right).
\end{equation}
To extract asymptotic behaviour for $T\ll h^2$ and $T\gg h^2$ we make
a crude saddle point, or ``pinching'' approximation, by assuming that
$\Phi(x)$ essentially vanishes for $|x|>1/2$ and is unity
otherwise. This fixes the random walker time $T$ via
$g_s-\mu(\Omega(T))=0$, while the number of terms in the summation is
restricted to satisfy $|g_s-\mu(\Omega(T))|\le
\sqrt{\sigma^2(\Omega(T))}$. After some algebra we find
\begin{subnumcases}{\elabel{regimes}\hspace*{-1cm}\pdf{g_s}{g_s;h}=}
\frac{h+1}{2}                     &\!\! for $g_s \ll 1/h$ \elabel{const_regime}\\
\frac{g^{-3/2}}{\sqrt{2\pi h}}    &\!\! for $1/h \ll g_s \ll 8 \pi h$ \elabel{intermediate_regime}\\
2 g^{-2}                          &\!\! for $g_s \gg 8 \pi h$ \elabel{large_regime}
\end{subnumcases}
\begin{figure}
\includegraphics[width=0.48\textwidth]{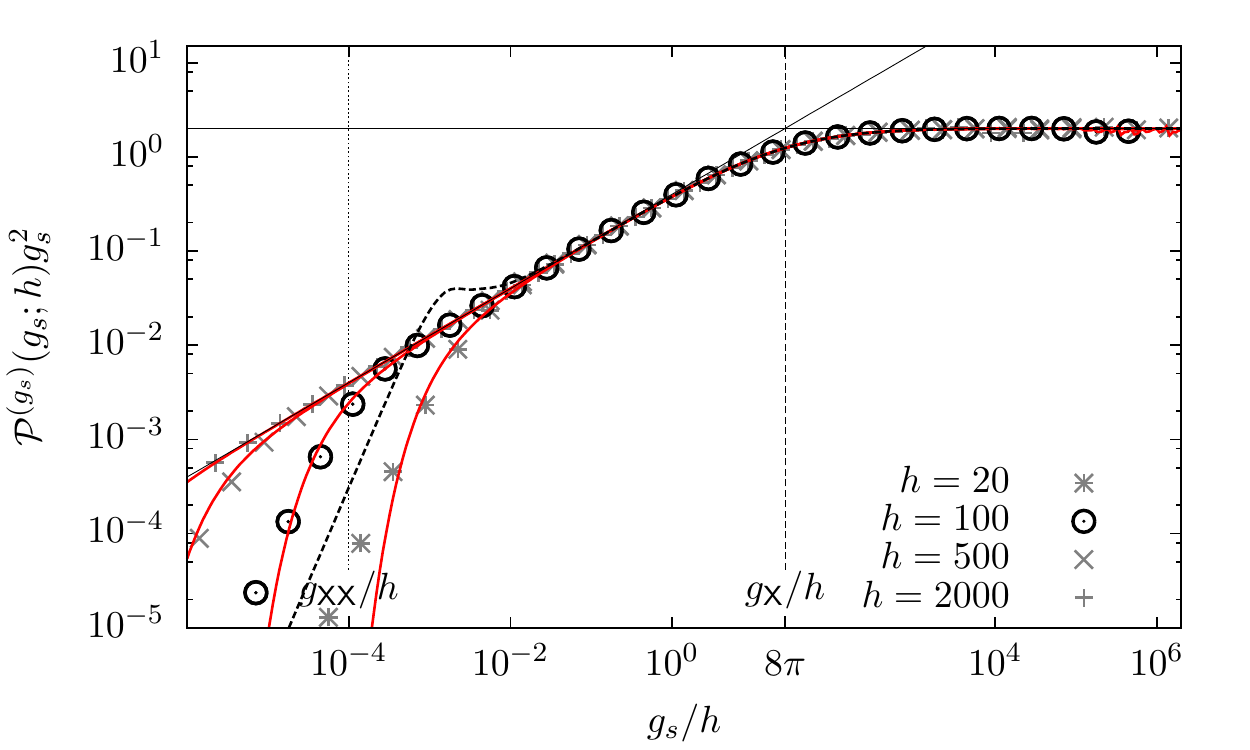}
\caption{\flabel{collapse} Collapse of the PDFs for different
  thresholds $h$ for large $g_s\gg1/h$, plotting
  $\pdf{g_s}{g_s;h}g_s^2$ against $g_s/h$, according to
  \Eref{handwaving_BD_PDF} and \Eref{large_collapse}, capturing
  \Eref{intermediate_regime} and \Eref{large_regime}.  The black full lines
  indicate the asymptotes according to \Eref{regimes}, the dashed lines
  show the crossovers at $g_s/h=8\pi$ and $g_s/h=1/h^2$ for
  $h=100$
  . Finally, the black thick dashed line corresponds to the analytical
 solution computed from \Eref{p_of_g} for $h=100$, while the red full lines were computed
 by numerically inverting the Laplace transform given in \Eref{laplace_trans}, 
 see \Aref{laplace_transform}. Another collapse is possible according to \Eref{small_collapse}.  }
\end{figure}

The qualitative scaling of these two asymptotes was anticipated after
\Eref{handwaving_BD_PDF}. The crossover time $g_{\cross}=8\pi h$,
shown in Figs.~\ref{fig:collapse} and \ref{fig:3regimes}, can be
determined by assuming continuity of $\pdf{g_s}{g_s;h}$ and thus
imposing $\frac{1}{\sqrt{2 h \pi}} g_{\cross}^{-3/2}=2
g_{\cross}^{-2}$.  \Fref{collapse} shows $\pdf{g_s}{g_s;h} g_s^2 $
versus $g_s/h$ for varying $h$, comparing Monte Carlo simulations for
varying $h$ with the numerical evaluation of \Eref{p_of_g} for
$h=100$, thus confirming the validity of the data collapse proposed in
\Eref{handwaving_BD_PDF}.  In particular, the shape of the transition
between the two asymptotic regimes, predicted to take place near
$g_\cross / h = 8 \pi$, is recovered from \Eref{p_of_g} with great
accuracy. As an alternative to the numerical evaluation of \Eref{p_of_g},
we introduce in \Aref{laplace_transform} a complementary approach that
provides the Laplace transform of $\pdf{g_s}{g_s;h}$, see
\Eref{laplace_trans}. Unfortunately, inverting the Laplace transform
analytically does not
seem feasible, but numerical inversion provides a perhaps simpler means of evaluating $\pdf{g_s}{g_s;h}$ in practice. 

\begin{figure}
\includegraphics[width=0.48\textwidth]{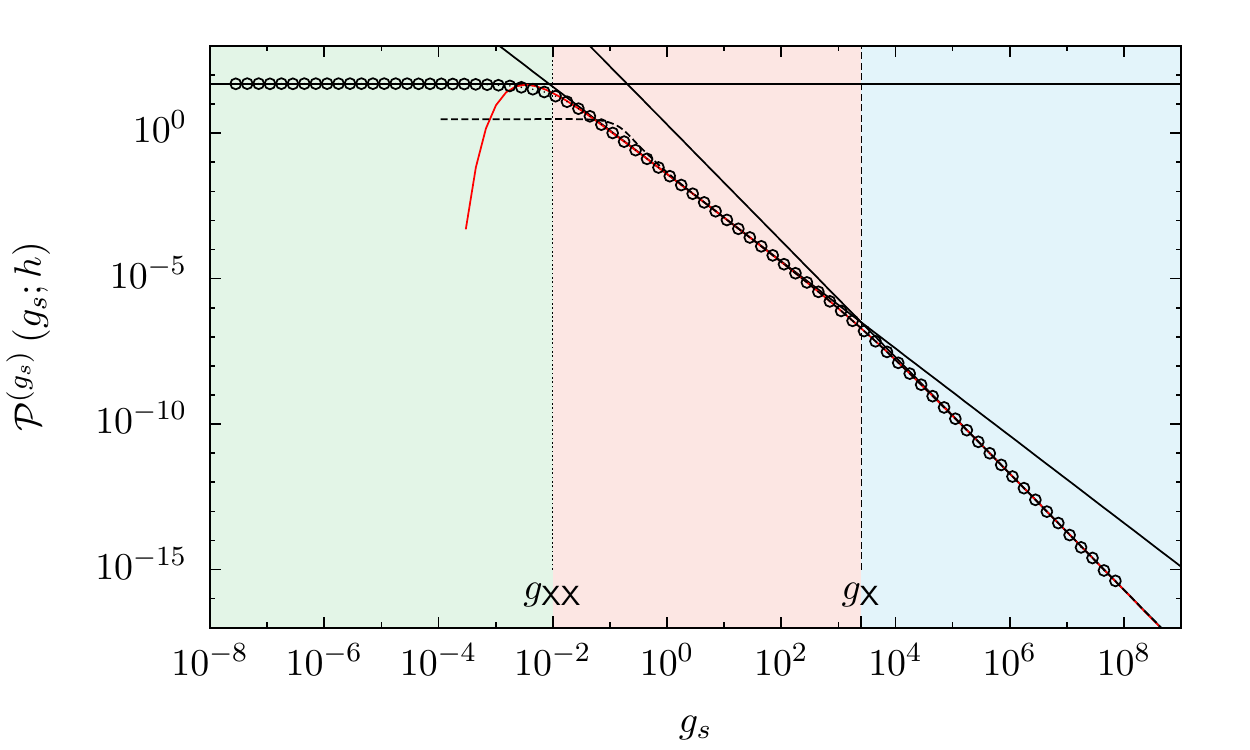}
\caption{\flabel{3regimes} The PDF of survival times
  $\pdf{g_s}{g_s;h}$ for $h=100$. Three scaling regimes partitioned by
  $g_\cross$ (thin dashed line) and $g_\ccross$ (thin dotted line) exist: For very
  short times $g_s \ll 1/h$ (green shading), the exponential waiting
  time to the first (death) event dominates, so that
  $\pdf{g_s}{g_s;h}\sim (h+1)/2$. For ``intermediate'' times (red
  shading) $1/h \ll g_s \ll 8 \pi h$, the effect of the threshold
  dominates, and hence $\pdf{g_s}{g_s;h} \sim g_s^{-3/2}/\sqrt{2 \pi
    h}$.  For long times (blue shading) $g_s \gg 8 \pi h$,
  $\pdf{g_s}{g_s;h}\sim 2g_s^{-2}$, independently of $h$.  Monte-Carlo
  simulation results are shown as symbols, asymptotes of
  $\pdf{g_s}{g_s;h}$, \Eref{regimes}, as solid lines, and the analytical solution
  $\pdf{g_s}{g_s;h}$, computed via \Eref{p_of_g} as a black thick dashed line, and via numerical inversion of the Laplace transform, \Eref{laplace_trans}, as a red solid line. 
  }
\end{figure}

In addition to the two asymptotic regimes discussed so far, one
notices that \fref{3regimes} displays yet another ``regime''
(left-most, green shading), which corresponds to extremely short
survival times. This regime is almost exclusively due to the walker
dying on the first move via the transition $n(0)=h+1$ to $n(g_s)=h$.
In this case, the sum in \Eref{g_traj} only has one term, and hence
the PDF of $g_s$ can be approximated as $\pdf{g_s}{g_s;h} =
\frac{1}{2}(h+1)\exp{-(h+1)g_s} \sim \frac{h+1}{2}$, where the factor
$1/2$ corresponds to the probability of $T=1$, and the limit of small
$g_s$ has been taken. Thus, for very short times $g_s \ll 1/h$, the
PDF of $g_s$ is essentially ``flat''. In order to estimate the
transition point to this third regime, we impose again continuity of
the solution, so that $(h+1)/2 = g_{\ccross}^{-3/2}/\sqrt{2 \pi h}$
and hence (dropping the constants) $g_{\ccross} = 1/h$, as shown in
\Eref{regimes} as well as Figs.~\ref{fig:collapse} and
\ref{fig:3regimes}.

Given the \emph{three} regimes shown in \Fref{collapse},
$\pdf{g_s}{g_s;h}$ can be collapsed either by ignoring the very short
scale, (see
\Eref{handwaving_BD_PDF})
\begin{equation}
\pdf{g_s}{g_s;h} \simeq 2 g_s^{-2} \GC_>(g_s/h) 
\qquad\text{for}\qquad
g\gg1/h
\elabel{large_collapse}
\end{equation}
with $\GC_>(x)=1$ for large $x$ and $\GC_>(x)=\sqrt{x/(8\pi)}$ in small $x$,
or according to
\begin{equation}
\pdf{g_s}{g_s;h} \simeq \frac{g_s^{-3/2}}{\sqrt{2\pi h}} \GC_<(g_s h) 
\qquad\text{for}\qquad
g\ll8\pi h
\elabel{small_collapse}
\end{equation}
with $\GC_<(x)=1$ for large $x$ and $\GC_<(x)=x^{3/2}\sqrt{\pi/2}$ for
small $x$. Power-law scaling (crossover) functions offer a number of
challenges, as they affect the ``apparent'' scaling exponent
\citep{ChristensenETAL:2008}. Also, there is no hard cutoff in the
present case, \ie moments $\ave{g_s^m}=\int\dint{g_s} \pdf{g_s}{g_s;h}
g_s^m$ do not exist for $m\ge2$.

\section{Discussion}\slabel{discussion}
The main goal of the present paper has been to understand how
thresholding influences data analysis. In particular, how thresholding
can change the scaling of observables and how one might detect this.

To this end, we worked through the consequences of thresholding in the
birth-death process, which is known to have a power-law PDF of
survival times with exponent $\gamma=2$.  We have shown, both
analytically and via simulations, that the survival times $g_s$ for
the thresholded process include a new scaling regime with exponent
$\gamma=3/2$ in the range $1/h \ll g_s \ll 8 \pi h$ (see
\Fref{3regimes}), where $h$ is the intensity level of the
threshold. 

We would like to emphasise how difficult it is to observe the
asymptotic $\gamma=2$ exponent, even for such an idealized toy
model. For large values of the threshold, $h=5\,000$, sample sizes as
large as $10^{10}$ are needed in order to populate the histogram for
large survival times. Real-world measurements are unlikely to meet the
demand for such vast amounts of data. An illustration of what might
then occur for realistic amounts of data that are subject to threshold
is given by \Fref{pdf_th_100_intro}, where only the threshold-induced
scaling regime associated with exponent $-3/2$ is visible.

Intriguingly, a qualitatively similar scaling phenomenology is
observed in renormalised renewal processes with diverging mean
interval sizes~\cite{Corral_jstat}. The random deletion of points
(that, together with a rescaling of time, constitutes the
renormalisation procedure) is analogous to the raising of a
threshold. It can be shown that the non-trivial fixed point
distribution of intervals is bi-power law. The asymptotic scaling
regime has the same exponent as that of the original interval
sizes. But, in addition, a prior scaling regime emerges with a
different exponent, and the crossover separating the two regimes moves
out with increasing threshold.

A fundamental difference between theoretical models and the analysis
of real-world processes is that in the former, asymptotic exponents
are defined in the limit of large events, with everything else
dismissed as irrelevant, whereas real world phenomena are usually
concerned with finite event sizes.  In our example, the effect of the
threshold dominates over the ``true'' process dynamics in the range
$1/h \ll g_s \ll 8 \pi h$, and grows with increasing $h$ before
eventually taking over the whole region of physical interest.
   
Of course, real data may not come from an underlying BDP. But we
believe that the specific lessons of the BDP apply more generally to
processes with multiplicative noise, \ie a noise whose amplitude
changes with the dynamical variable (the degree of freedom): At large
thresholds small changes of that variable are negligible and an
effectively additive process is obtained (the plain random walker
above). Only for large values of the dynamical variable is the
original process recovered. These large values are rare, in
particular when another cutoff (such as, effectively, the sample size)
limits the effective observation time ($2T-1$ above). In this context
it is worth mentioning the work of Laurson \etal \cite{Laurson_upon},
in which thresholds were applied to Brownian excursions. However,
since noise is additive in Brownian motion, the effect of thresholding
is relatively benign. Indeed, no new scaling regime appears as a
result of thresholding, and the asymptotic exponent of $-3/2$ is
recovered no matter what threshold is applied.

In the worst case, thresholding may therefore bury the asymptotics
which would only be recovered for \emph{much} longer observation
times. However, if the threshold can easily be changed, its effect can
be studied systematically by attempting a data collapse onto the
scaling ansatz $\pdf{g_s}{g_s;h} = g_s^{-\gamma} \GC(g_s/h^D)$,
Eqs.~\eref{handwaving_BD_PDF} and \eref{large_collapse}, with
exponents $\gamma$ and $D$ to be determined, as performed in
\Fref{collapse} with $\gamma=2$ and $D=1$. The threshold plays an
analogous r\^ole to the system size in finite-size scaling (albeit for
intermediate scales). In the present case, the exponents in the
collapse, together with the asymptote of the scaling function,
identify two processes at work, namely the BDP as well as the random
walk.

\begin{widetext}
\appendix
\section{Mean and variance of the survival time}\slabel{mean_var_appendix}
This appendix contains the details of the calculations leading to the
approximation (in large $T$), 
\Eref{mu_Omega_largeT_mainText} and \Eref{sigma2_in_IKmu_mainText}, as
well as their asymptotes \Eref{mu_asympt_mainText} and
\Eref{sigma2_asympt_mainText},
for the mean $\mu(\Omega)$ and the variance
$\sigma^2(\Omega)$ respectively, averaged
over the ensemble $\Omega(T)$, or $\Omega$ for short, of the mapped random walks with the constraint that
they terminate at $2T-1$, see \Fref{rho_illustration}.

\begin{figure}
\resizebox{0.49\textwidth}{!}{
\includegraphics{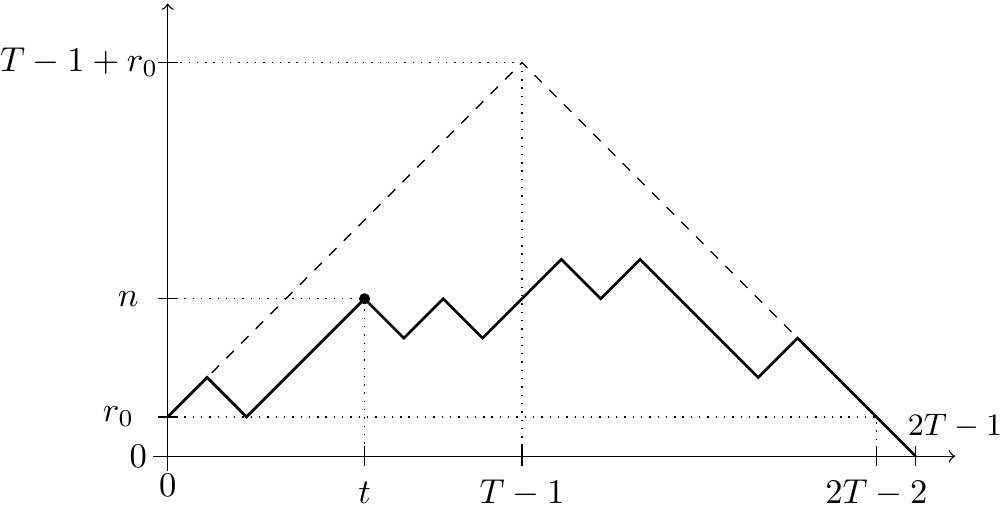}
}
\caption{\flabel{rho_illustration} Sample path of a random walk along an
absorbing wall at $0$. The walker starts at $t=0$ from $r_0$ and
terminates at $2T-1$ by reaching the wall $r_{2T-1}=0$, \ie
$r_{2T-2}=1$. By construction, it cannot escape from the region
demarcated by the dashed line. When counting distinct paths, the number
of paths terminating at $r_{2T-1}=0$ equals the number of paths passing
through $r_{2T-2}=1$.}
\end{figure}

In the following, we will use the notation $\xi_t$ for $\xi_t(r_t+h)$,
but it is important to note that any two $\xi_t(r_t+h)$ are
independent, even though the consecutive $r_t$ are not. The random
variable $g_s(\RC)$ in \Eref{g_traj} is thus a sum of
\emph{independent} random variables $\xi_t$, whose mean and variance
at consecutive $t$, however, are correlated due to $r_t$ being a
trajectory of a random walk.  Because $h+r_t>0$ for $t<2T-1$, the
limiting distribution of $(g_s(\RC)-\mu(\RC))/\sqrt{\sigma^2(\RC)}$ as
$T\to\infty$ is a Gaussian with unit variance. Mean $\mu(\RC)$ and
variance $\sigma^2(\RC)$ are defined as
\begin{subeqnarray}{\elabel{mu_sigma_explicitly}}
\mu(\RC)&=&\ave{g_s(\RC)}_{\RC}=\sum_{t=0}^{2T-2} \ave{\xi_t}_{\RC} \\
\sigma^2(\RC)&=&\ave{(g_s(\RC))^2}_{\RC}-\ave{g_s(\RC)}^2_{\RC} \nonumber \\
&=&\sum_{t,t'=0}^{2T-2} \ave{\xi_t\xi_{t'}}_{\RC}
-  \ave{\xi_t}_{\RC}\ave{\xi_{t'}}_{\RC}
\end{subeqnarray}
and are functions 
of the trajectory $\RC$ with $\ave{\cdot}_{\RC}$ taking the expectation
across the ensemble of $\xi$ for given, fixed $\RC$, \ie $\ave{\xi_t}_{\RC}=1/(r_t+h)$ and 
$\ave{\xi_t^2}_{\RC}-\ave{\xi_t}^2_{\RC}=1/(r_t+h)^2$. Because
$\ave{\xi_t\xi_{t'}}_{\RC}=\ave{\xi_t}_{\RC}\ave{\xi_{t'}}_{\RC}$
for $t\ne t'$ the mean and the variance are in fact just
\begin{subeqnarray}{\elabel{mu_sigma_with_Poisson_in}}
\mu(\RC)&=&\sum_{t=0}^{2T-2} \frac{1}{r_t+h} \\
\sigma^2(\RC)&=& \sum_{t=0}^{2T-2} \frac{1}{(r_t+h)^2} \ .
\end{subeqnarray}
If $\rho_n(\RC)$ \emph{counts} the number of times $r_t$ attains a
certain level,
\begin{equation}
\rho_n(\RC)=\sum_{t=0}^{2T-2} \delta_{n,r_t} 
\elabel{def_rho}
\end{equation}
then $\sum_{t=0}^{2T-2} f(r_t)=\sum_{t=0}^{2T-2} \sum_{n=0}^{\infty}
\delta_{n,r_t} f(n) = \sum_{n=0}^{\infty} \rho_n(\RC) f(n)$, so
\begin{subeqnarray}{}
\mu(\RC)&=&\sum_{n=r_0}^{T-1+r_0} \frac{\rho_n(\RC)}{n+h} \\
\elabel{sigma_fixed_R}
\sigma^2(\RC)&=&\sum_{n=r_0}^{T-1+r_0} \frac{\rho_n(\RC)}{(n+h)^2} \ .\\
\end{subeqnarray}
where we used the fact that within time $2T-2$ our random walker cannot stray further away
from $r_0$ than $T-1+r_0$, as illustrated in \Fref{rho_illustration}.

In the same vein, we can now proceed to find mean and variance of
$g_s$ over the entire ensemble $\Omega=\Omega(T)$ of trajectories $\RC$ that
terminate at $2T-1$. In the following $\ave{\cdot}_{\Omega}$ denotes the ensemble
average over all trajectories $\RC\in\Omega$, each appearing with the
same probability
\begin{equation}
\ave{f(\xi_t)}_{\Omega} = \frac{1}{|\Omega|} \sum_{\RC}
\ave{f(\xi_t)}_{\RC}
\end{equation}
where $f(\xi_t)$ is an arbitrary function of the random variable
$\xi_t$. We therefore have
\begin{equation}\elabel{first_moment_full}
\mu(\Omega)=
 \ave{\sum_{t=0}^{2T-2} \xi_t}_{\Omega}
=\frac{1}{|\Omega|} \sum_{\RC} \sum_{t=0}^{2T-2} \frac{1}{r_t+h}
=\frac{1}{|\Omega|} \sum_{\RC} \sum_{n=r_0}^{T-1+r_0} \frac{\rho_n(\RC)}{n+h}
=\sum_{n=r_0}^{T-1+r_0} \frac{\ave{\rho_n(\RC)}_{\Omega}}{n+h}
\end{equation}
where $\ave{\rho_n(\RC)}_{\Omega}$ is in fact the expected number of times a
random walker terminating at $2T-1$ attains level $n$. 

The variance turns out to require a bit more work. The second moment
\begin{equation}
\ave{g_s(\RC)^2}_{\Omega}
= \ave{\left(\sum_{t=0}^{2T-2} \xi_t\right)^2}_{\Omega}
= \frac{1}{|\Omega|} \sum_{\RC} \sum_{t,t'=0}^{2T-2} 
\ave{\xi_t \xi_{t'}}_{\RC}
\end{equation}
simplifies significantly when $t\ne t'$ in which case the lack of
correlations means that the expectation factorises
$\ave{\xi_t \xi_{t'}}_{\RC}=\ave{\xi_t}_{\RC} \ave{\xi_{t'}}_{\RC}$, so
that we can write 
\begin{equation}
\sum_{t,t'=0}^{2T-2} \ave{\xi_t \xi_{t'}}_{\RC}
= \sum_{t,t'=0}^{2T-2} \ave{\xi_t }_{\RC} \ave{\xi_{t'} }_{\RC}
+ \sum_{t=0}^{2T-2} \left( 
\ave{\xi_t^2}_{\RC} - \ave{\xi_t}^2_{\RC}
\right)
\end{equation}
Obviously $\sum_{t,t'=0}^{2T-2} \ave{\xi_t }_{\RC} \ave{\xi_{t'} }_{\RC}
=\left(\sum_{t=0}^{2T-2} \ave{\xi_t }_{\RC}\right)^2$, but that is not a useful simplification for the
time being.

The square of the first moment, \Eref{first_moment_full}, is best
written as
\begin{equation}
\ave{g_s(\RC)}^2_{\Omega} 
= \frac{1}{|\Omega|^2} \sum_{\RC,\RC'} \sum_{t,t'=0}^{2T-2}
\ave{\xi_t}_{\RC}
\ave{\xi_{t'}}_{\RC'}
\end{equation}
so that
\begin{multline}
\sigma^2(\Omega)=\ave{g_s(\RC)^2}_{\Omega}-\ave{g_s(\RC)}^2_{\Omega}
\\
= 
\frac{1}{|\Omega|} \sum_{\RC} 
\sum_{t,t'=0}^{2T-2} \ave{\xi_t }_{\RC} \ave{\xi_{t'} }_{\RC}
+ \frac{1}{|\Omega|} \sum_{\RC} 
\sum_{t=0}^{2T-2} \left( \ave{\xi_t^2}_{\RC} - \ave{\xi_t}^2_{\RC} \right)
- \frac{1}{|\Omega|^2} \sum_{\RC,\RC'} 
\sum_{t,t'=0}^{2T-2} \ave{\xi_t}_{\RC} \ave{\xi_{t'}}_{\RC'}
\ .
\end{multline}
The first and the last pair of sums can be written as
\begin{equation}
\frac{1}{|\Omega|^2} \sum_{\RC,\RC'}
\sum_{t,t'=0}^{2T-2} \ave{\xi_t}_{\RC} \Big(
\ave{\xi_{t'}}_{\RC} - \ave{\xi_{t'}}_{\RC'} 
\Big)
\end{equation}
using $\sum_{\RC}(1/|\Omega|)=1$, so that
\begin{equation}
\sigma^2(\Omega)=
\frac{1}{|\Omega|^2} \sum_{\RC,\RC'} 
\sum_{t,t'=0}^{2T-2} \ave{\xi_t}_{\RC} \Big( \ave{\xi_{t'}}_{\RC} - \ave{\xi_{t'}}_{\RC'} \Big)
+
\frac{1}{|\Omega|} \sum_{\RC}
\sum_{t=0}^{2T-2} \Big( \ave{\xi_t^2}_{\RC} - \ave{\xi_t}^2_{\RC} \Big)
\end{equation}
In the first sum, the two terms can be separated into those in $t'$ and one in $t$.
Using the same notation as above, \Eref{def_rho} we have
\begin{equation}
\sum_{t'=0}^{2T-2} \Big( \ave{\xi_{t'}}_{\RC} - \ave{\xi_{t'}}_{\RC'} \Big) =
\sum_{n'=r_0}^{T-1+r_0} \frac{\rho_{n'}(\RC)-\rho_{n'}(\RC')}{n'+h}
\end{equation}
and $\sum_{t=0}^{2T-2} \ave{\xi_t}_{\RC} = \sum_{n=r_0}^{T-1+r_0} \frac{\rho_n(\RC)}{n+h}$. 

The second sum recovers the earlier result in \Eref{sigma_fixed_R}, as 
$\ave{\xi_t^2}_{\RC}=\frac{2}{(r_t+h)^2}$ and $\ave{\xi_t}_{\RC}=\frac{1}{r_t+h}$, 
so that
\begin{equation}
\sum_{t=0}^{2T-2} \left( \ave{\xi_t^2}_{\RC} - \ave{\xi_t}^2_{\RC} \right)
=
\sum_{n=r_0}^{T-1+r_0} \frac{\rho_n(\RC)}{(n+h)^2}
\end{equation}
and therefore
\begin{multline}\elabel{variance_full}
\sigma^2(\Omega)=
\frac{1}{|\Omega|^2} \sum_{\RC,\RC'} 
\sum_{n,n'=r_0}^{T-1+r_0} \frac{\rho_n(\RC)}{n+h} \frac{\rho_{n'}(\RC)-\rho_{n'}(\RC')}{n'+h}
+
\frac{1}{|\Omega|} \sum_{\RC}
\sum_{n=r_0}^{T-1+r_0} \frac{\rho_n(\RC)}{(n+h)^2}\\
= 
  \frac{1}{|\Omega|} \sum_{\RC} \sum_{n,n'=r_0}^{T-1+r_0} \frac{\rho_n(\RC)\rho_{n'}(\RC)}{(n+h)(n'+h)}
- \left(\frac{1}{|\Omega|} \sum_{\RC} \sum_{n=r_0}^{T-1+r_0} \frac{\rho_n(\RC)}{n+h}\right)^2
+
\frac{1}{|\Omega|} \sum_{\RC}
\sum_{n=r_0}^{T-1+r_0} \frac{\rho_n(\RC)}{(n+h)^2}\\
= 
  \sum_{n,n'=r_0}^{T-1+r_0} \frac{\ave{\rho_n(\RC)\rho_{n'}(\RC)}_{\Omega}}{(n+h)(n'+h)}
- \left(\sum_{n=r_0}^{T-1+r_0} \frac{\ave{\rho_n(\RC)}_{\Omega}}{n+h}\right)^2
+
\sum_{n=r_0}^{T-1+r_0} \frac{\ave{\rho_n(\RC)}_{\Omega}}{(n+h)^2}
\end{multline}
We now have the mean $\mu(\Omega)$, \Eref{first_moment_full}, 
and the variance $\sigma^2(\Omega)$, \Eref{variance_full}, in terms of
$\ave{\rho_n(\RC)}_{\Omega}$ and
$\ave{\rho_n(\RC)\rho_{n'}(\RC)}_{\Omega}$. In the following, we will
determine these two quantities and then return to the original task of
finding a closed-form expression for $\mu(\Omega)$ and
$\sigma^2(\Omega)$.

\subsubsection{$\ave{\rho_n(\RC)}_{\Omega}$ and
$\ave{\rho_n(\RC)\rho_{n'}(\RC)}_{\Omega}$}
Of the 
two expectations, $\ave{\rho_n(\RC)}_{\Omega}$ is obviously the easier
one to determine. In fact, $\sum_n\rho_n(\RC)=2T-1$ implies
$\sum_{n'}\ave{\rho_n(\RC)\rho_{n'}(\RC)}=(2T-1)\ave{\rho_n(\RC)}$,
\ie $\ave{\rho_n(\RC)}_{\Omega}$ is a ``marginal'' of
$\ave{\rho_n(\RC)\rho_{n'}(\RC)}_{\Omega}$.

\newcommand{\binomial}[2]{{#1 \choose #2}}

To determine $\ave{\rho_n(\RC)}_{\Omega}$, we use the method of images
(or mirror charges).
The number of positive paths ($r_i>0$) from $(t=0,r_0)$ to $(t,n)$
are $\binomial{t}{\frac{n-r_0+t}{2}}-\binomial{t}{\frac{n+r_0+t}{2}}$
for $n+r_0+t$ even and $n>0$. 
By construction, the number of paths passing
through $n=0$ is exactly $0$, thereby implementing the boundary
condition. 
The set of paths (to be considered in the following)
which terminate at time $2T-1$ by reaching $r_{2T-1}=0$ is, up to the
final step, identical to
the set of paths passing through $(2T-2,1)$, \ie $r_{2T-1}=0$.
The number of positive paths (see
\Fref{rho_illustration}) originating
from $(0,r_0=1)$ and terminating at $(t=2T-1,r_{2T-1}=0)$ therefore equals the
number of positive paths from $(0,r_0=1)$ to $(t=2T-2,n=1)$, so that
$|\Omega|=\binomial{2T-2}{T-1}-\binomial{2T-2}{T}=\frac{1}{T}\binomial{2T-2}{T-1}$,
which are the Catalan numbers \cite{Knuth:1997V1,Stanley:1999}. For $r_0=1$ we also have
\begin{equation}
\binomial{t}{\frac{n-1+t}{2}}-\binomial{t}{\frac{n+1+t}{2}}
=
\frac{n}{t+1}
\binomial{t+1}{\frac{n+1+t}{2}}
\end{equation}
again for $n+r_0+t$ even. This is the number of positive paths from $(0,1)$ to
$(t,n)$ and by symmetry also the number of paths from $(2T-2-t,n)$ to
$(2T-2,1)$, given that the walk is unbiased (see
\Fref{rho_illustration}). 
If $\ave{\rho_n(t;\RC)}_{\Omega}$ is the expected fraction of paths
passing through $(t,n)$ (illustrated in \Fref{rho_illustration}), we therefore have 
\begin{equation}\elabel{rho_n_carefully}
\ave{\rho_n(t;\RC)}_{\Omega} =
\underbrace{\frac{T}{\binomial{2T-2}{T-1}}}_{1/|\Omega|}
\underbrace{\frac{n}{t+1}
\binomial{t+1}{\frac{n+1+t}{2}}}_{\text{from $(0,1)$ to $(t,n)$}}
\underbrace{\frac{n}{2T-1-t}
\binomial{2T-1-t}{\frac{n+ 2T-1-t}{2}}}_{\text{from $(t,n)$ to
$(2T-2,1)$}}
\end{equation}
which is normalised by construction, \ie $\sum_n
\ave{\rho_n(t;\RC)}_{\Omega} =1$. The first binomial factor in the
denominator is due to the normalisation, whereas of the last two, the
first is due to paths from $(0,1)$ to $(t,n)$ and the second due to
paths from $(t,n)$ to $(2T-2-t,1)$.
In the following we are interested in
the fraction of times a random walker reaches a certain level during its
lifetime, $\ave{\rho_n(\RC)}_{\Omega} = \sum_t
\ave{\rho_n(t;\RC)}_{\Omega}$. Using 
$
\binom{a}{b} \simeq
2^a 
({a \pi}/{2})^{-1/2}
\exp{ -\frac{2}{a}\left(  b -  \frac{a}{2}\right)^2 }
$
we find
\newcommand{\ttilde}{\tilde{t}}
\begin{equation}
\ave{\rho_n(t;\RC)}_{\Omega} \simeq
\frac{8T^{3/2}}{\sqrt{\pi}}
\frac{n^2}{\ttilde^{3/2}(2T-\ttilde)^{3/2}}
\Exp{-\frac{n^2}{2\ttilde}-\frac{n^2}{2(2T-\ttilde)}}
\ ,
\end{equation}
where we have used $T\gg1$ and $\ttilde=t+1$. Simplifying
further gives
\begin{equation}
\ave{\rho_n(\RC)}_{\Omega} =
\sum_{\ttilde=n}^{2T-n} \ave{\rho_n(t;\RC)}_{\Omega}
\simeq
8 \nu^2 \sqrt{\frac{T}{\pi}}
\sum_{\ttilde=n}^{2T-n}
\frac{\exp{-\frac{\nu^2}{\tau(2-\tau)}}}{T (\tau(2-\tau)^{3/2}}
\end{equation}
with the sum running over the $\ttilde$ with the correct parity and 
$\tau=\ttilde/T$ and $\nu=n/\sqrt{T}$. In the limit of large $T\gg1$ we
find \cite{Mathematica:8.0.1.0}
\begin{equation}
\lim_{T\to\infty}\frac{\ave{\rho_n(\RC)}_{\Omega}}{\sqrt{T}}=
\frac{4\nu^2}{\sqrt{\pi}} 
\int_0^2\dint{\tau}
\frac{\exp{-\frac{\nu^2}{\tau(2-\tau)}}}{(\tau(2-\tau))^{3/2}}
=
4 \nu \exp{-\nu^2}
\end{equation}
where the parity has been accounted for by dividing by $2$. In the last
step, the integral was
performed by some substitutions, as 
$\tau(2-\tau)$ is symmetric about $1$.
It follows that in the limit of large $T\gg1$
\begin{equation}\elabel{rho_Omega_largeT}
\ave{\rho_n(\RC)}_{\Omega}
\simeq
4 n 
\exp{-\frac{n^2}{T}}
\end{equation}
Using that expression in \Eref{first_moment_full} gives
\Eref{mu_Omega_largeT_mainText}, namely
\begin{equation}\elabel{mu_Omega_largeT}
\frac{\mu(\Omega)}{\sqrt{T}} \simeq 4
\sum_{n=r_0}^{T-1+r_0} \frac{1}{\sqrt{T}} \frac{\nu}{\nu+\frac{h}{\sqrt{T}}}
\exp{-\nu^2}
\simeq
\int_0^{\sqrt{T}} \dint{\nu} \frac{4\nu}{\nu+\frac{h}{\sqrt{T}}}
\exp{-\nu^2}
\simeq
\int_0^{\infty} \dint{\nu} \frac{4\nu}{\nu+\frac{h}{\sqrt{T}}}
\exp{-\nu^2}
=
2 \sqrt{\pi} +
2 \frac{h}{\sqrt{T}}
\psi\left( \frac{h}{\sqrt{T}} \right)
\end{equation}
with 
\citep[][Eq.~27.6.3]{AbramowitzStegun:1970}
\begin{equation}
\psi(x) = -  \exp{-x^2} 
\left( 
 2 \sqrt{\pi} \int_0^x\dint{s}\exp{s^2}
 +
\dashint_{-x^2}^\infty \dint{y} \frac{\exp{-y}}{y}
\right)
\elabel{def_psi}
\end{equation}
where we have used $r_0=1$. 
The first integral is known as the exponential integral
function $\dashint_{-x}^\infty \dint{y} \frac{\exp{-y}}{y}=-\operatorname{Ei}(x)$ 
and the second as (a multiple of) the imaginary error function $2
\sqrt{\pi} \int_0^x\dint{s}\exp{s^2}=\pi \EC(\imag x)/\imag$.
In the limit of large arguments $x$, the function $\psi(x)$
is $-\sqrt{\pi}/x+1/x^2-\sqrt{\pi}/(2x^3)+1/x^4+\OC(x^{-5})$, in the limit of small arguments by
$\gamma+2\ln(x)$, where $\gamma$ is the Euler-Mascheroni constant. We
conclude that 
\begin{subnumcases}{\mu(\Omega)\simeq\elabel{mu_asympt}}
2\sqrt{\pi T} + 2 h (\gamma+2\ln(h/\sqrt{T})) & for $T\gg h^2$ \\
2 T/h - \sqrt{\pi} T^{3/2}/h^2 + 2 T^2/h^3 & for $T\ll h^2$
\end{subnumcases}
(see \Eref{mu_asympt_mainText})
provided $T$ is large compared to $1$, which is the key assumption of
the approximations used above. It is worth stressing this distinction:
$T$ has to be large compared to $1$ in order to make the various
continuum approximations (effectively continuous in time, so sums turn
into integrals and continuous in state, so binomials can be approximated
by Gaussians), but no restrictions were made regarding
the ratio $T/h^2$.

The correlation function $\ave{\rho_n(\RC)\rho_{n'}(\RC)}_{\Omega}$ can
be determined using the same methods, starting with \Eref{rho_n_carefully}:
\begin{align}
\elabel{rho_nm}
\ave{\rho_n(t;\RC) \rho_{n'}(t';\RC)}_{\Omega}
&= \sum_{t} \sum_{t'<t}
\underbrace{\frac{T}{\binomial{2T-2}{T-1}}}_{1/|\Omega|}
\underbrace{\frac{n}{t'+1}\binom{t'+1}{\frac{n+t'+1}{2}}
}_{\text{from $(0,1)$ to $(t',n)$}}
\biggl[
\underbrace{
	\binom{t-t'}{\frac{t-t'+n-n'}{2}}-\binom{t-t'}{\frac{t-t'+n+n'}{2}}
}_{\text{from $(t',n)$ to $(t,n')$}}
\biggr]
\underbrace{\frac{n'}{2T-1-t} \binom{2T-1-t}{\frac{n'+2T-1-t}{2}}
}_{\text{from $(t,n')$ to $(2T-2,1)$}}\\
&+ \sum_{t} \sum_{t'\ge t}
\underbrace{\frac{T}{\binomial{2T-2}{T-1}}}_{1/|\Omega|}
\underbrace{\frac{n}{t+1}\binom{t+1}{\frac{n+t+1}{2}}
}_{\text{from $(0,1)$ to $(t,n)$}}
\biggl[
\underbrace{
	\binom{t'-t}{\frac{t'-t+n-n'}{2}}-\binom{t'-t}{\frac{t'-t+n+n'}{2}}
}_{\text{from $(t,n)$ to $(t',n')$}}
\biggr]
\underbrace{\frac{n'}{2T-1-t} \binom{2T-1-t}{\frac{n'+2T-1-t'}{2}}
}_{\text{from $(t',n')$ to $(2T-2,1)$}}\nonumber
\end{align}
Because both $t$ and $t'$ are dummy variables, one might be tempted to
write the entire expression as twice the first double sum, which is
indeed correct as long as $n\ne n'$. In that case, the case $t'=t$ does
not contribute because the ``middle chunk'' (from  $(t,n)$ to $(t',n')$)
vanishes. However, if $n=n'$ that middle chunk is unity and therefore
needs to be included separately. This precaution turns out to be
unnecessary once the binomials are approximated by Gaussians and the
sums by integrals.

The resulting convolutions are technically
tedious, but can be determined in closed form on the basis of Laplace
transforms and tables 
\citep[][Eq.~29.3.82 and Eq.~29.3.84]{AbramowitzStegun:1970}, 
resulting finally in 
\begin{equation}\elabel{rhorho_Omega_largeT}
\ave{\rho_n (\RC) \rho_{n'}(\RC)}_{\Omega}
\simeq 8 T(\exp{- n^2/T } - \exp{-(n+n')^{2}/T})
\end{equation}
to leading order in $T$. 

We proceed to determine 
\Eref{variance_full} using \Eref{rho_Omega_largeT} and
\Eref{rhorho_Omega_largeT} in the limit of large $T$. Again, we
interpret the sums as Riemann sums, to be approximated by integrals,
resulting in \Eref{sigma2_in_IKmu_mainText},
\begin{equation}
\sigma^2(\Omega) \simeq  T\ \IC(x) - \mu(\Omega)^2 + \KC(x)
\elabel{sigma2_in_IKmu}
\end{equation}
with $x=h/\sqrt{T}$ and
\begin{subeqnarray}{\elabel{KI_integrals}}
\KC(x) & = & \int_0^\infty \dint{n} \frac{4n \exp{-n^2}}{(n+x)^2} = -4 + 4 x \sqrt{\pi} +2(2x^2-1)\psi(x)\\
\IC(x) & = & 16 \int_0^\infty \dint{n} \int_0^n \dint{n'} \frac{ \exp{-n^2}-\exp{-(n+n')^2}}{(n+x)(n'+x)}
\end{subeqnarray}
(for the definition of $\psi(x)$ see \Eref{def_psi}). Unfortunately, we
were not able to reduce $\IC(x)$ further. 

Because of the structure of \Eref{sigma2_in_IKmu}, where 
$T\ \IC(x) r - \mu(\Omega)^2$ scale linearly in $T$ at fixed $x=h/\sqrt{T}$, whereas $\KC(x)$
remains constant, a statement about the leading order behaviour in $T$
is no longer equivalent to a statement about the leading order behaviour
in $1/x^2$. 
This is complicated
further by the assumption made throughout that $T$ is large. The limits
we are interested in, are in fact $T\gg h^2$ with $T\gg1$ and $1\ll T\ll h^2$.
In the following, we need to distinguish not only large $x$ from small
$x$, but also different orders of $T$.

It is straightforward to
determine the asymptote of $\IC(x)$ in large $x$, where the denominator of the integrand is dominated
by $x^2$ while the numerator vanishes at least as fast as $\exp{-n^2}$,
because
$\exp{-n^2}-\exp{-(n+n')^2}=\exp{-n^2}(1-\exp{-2nn'-n'^2}$ and
$0\le(1-\exp{-2nn'-n'^2})<1$, so \citep{Mathematica:8.0.1.0}
\begin{equation}
\IC(x) = 
\frac{16}{x^2} \int_0^\infty \dint{n}\int_0^n\dint{n'}
\exp{-n^2}(1-\exp{-2nn'-n'^2})
\left(
1-\frac{n}{x}+\frac{n^2}{x^2}+\ldots
\right)
\left(
1-\frac{n'}{x}+\frac{n'^2}{x^2}+\ldots
\right)
= 
\frac{4}{x^2} - \frac{4\sqrt{\pi}}{x^3}+\frac{34}{3x^4}+\OC(x^{-5})
\end{equation}
Similarly, or using the expansion of
$\psi(x)$ introduced above, we find $\KC(x)=2/x^2 + \OC(x^{-3})$. 
Since $\mu(\Omega)=T(2/x-\sqrt{\pi}/x^2+2/x^3+\ldots)$, the first two
terms in the expansion of $\IC(x)$ for large $x$ cancel, and we
arrive at 
\begin{equation}\elabel{sigma2_expansion}
\sigma^2(\Omega)=\frac{2}{x^2}+\OC(x^3) +
T\left(\frac{34}{3x^4}-\frac{8+\pi}{x^4}+\OC(x^5)\right)
=\frac{2T}{h^2}+\frac{10-3\pi}{h^4}T^3+\ldots
\end{equation}
for $T\ll h^2$, containing the rather unusual looking (``barely
positive'', one might say)
difference $10-3\pi$. The second term in \Eref{sigma2_expansion} is clearly subleading
in large $x$ and no ambiguity arises in that limit, not even if $T\gg1$.

The limit $h/\sqrt{T} = x\to0$, on the other hand, $\IC(x)$ is 
\begin{equation}
\IC(x) = \frac{4}{3}\pi^2 + \OC(x)
\end{equation}
using
\citep[][Eq.~27.7.6]{AbramowitzStegun:1970}
so that $T\IC(x)-\mu(\Omega)^2=T(4\pi^2/3-4\pi+\OC(x))$, whereas
$\KC(x)=-4\ln(x)-4-2\gamma$ \emph{diverges} in small $x$. Although
this latter term therefore dominates in small $x$, the former,
$T\IC(x)-\mu(\Omega)^2$, does 
for large $T\gg h^2$ at
finite, fixed $h$. 

We are now in the position to determine the relevant asymptotes of
$\sigma^2(\Omega)$, as stated in \Eref{sigma2_asympt_mainText},
\begin{subnumcases}{\sigma^2(\Omega)\simeq\elabel{sigma2_asympt}}
4 \pi T \frac{\pi-3}{3} & for $T\gg h^2$ \\
\frac{2T}{h^2}  & for $T\ll h^2$
\end{subnumcases}

\section{Limiting distribution of $g_s(\Omega)/\sqrt{T}$}\slabel{limiting_dist}

In this second appendix, we explicitly find the limiting distribution
of $g_s(\Omega)/\sqrt{T}$. We begin by noting that, for $T\gg1$,
$g_s(\Omega)$ can be approximated as $ g_s(\Omega) \simeq
\int_0^{2T}dt \frac{1}{x(t)+h} $, where $x(t)$ performs a Brownian
excursion of length $2T$. While for large but finite $T$ this is
clearly an approximation (\eg the exponential random variables have
been replaced by their mean), in the limit of $T\to \infty$ the
approximation becomes exact. In particular, the ``noise'' due to the
variance of the exponential random variables scales like $\log T $,
see \Eref{KI_integrals}, and thus vanishes after rescaling with
respect to $\sqrt{T}$. In addition, owing to the scaling properties of
Brownian motion,
\begin{equation}
\lim_{T\to\infty}g_s(\Omega)/\sqrt{T} = \lim_{T\to\infty} \int_0^2 dt \frac{1}{x(t)+h/\sqrt{T}} = \int_0^2 dt \frac{1}{x(t)}
\end{equation}
where $x(t)$ is a Brownian excursion of length 2. To find the
distribution of this quantity, we first define $y(t)=\int_0^t dt'
1/x(t')$, and the propagator $Z(x,y,x_0,y_0,t)$, \ie the probability
for a Brownian particle to go from $(x_0,y_0)$ to $(x,y)$ in time $t$,
without touching the line $x=0$. Using standard techniques
\cite{Chaichian:2001}, the associated Fokker-Plank equation for the
propagator takes the form
\begin{equation}
\left[
\partial_{t}
+ \frac{1}{x} \partial_{y}
-\frac{1}{2}\partial_{xx}
\right] Z(x,y,x_0,y_0,t) = 0,
\end{equation}
with initial condition
\begin{equation}
Z(x,y,x_0,y_0,0) = \delta(x-x_0) \delta(y-y_0),
\end{equation}
and boundary condition
\begin{equation}
Z(0,y,x_0,y_0,t) = 0.
\end{equation}
Taking the Laplace transform with respect to $t$ yields
\begin{alignat}{1}
\left[
s
+ \frac{1}{x} \partial_{y}
-\frac{1}{2}\partial_{xx}
\right] \hat{Z}(x,y,x_0,s)&= \delta(x-x_0) \delta(y) \\
\hat{Z}(0,y,x_0,s) &= 0
\end{alignat}
We first solve the associated homogeneous equation, from which we will
be able to construct the solution to the inhomogeneous problem. After
substituting the ansatz
$\hat{Z}_\text{hom}(x,y,s)=\Psi(x,s)\rho(y,s)$, the equation separates
into
\begin{alignat}{1}
\label{eigprob}
-1/2 \partial_{xx}\Psi(x,s) + (s-\lambda/x)\Psi(x,s) &= 0 \\
-\partial_y \rho(y,s) + \lambda \rho(y,s) &=0,
\end{alignat}
where $\lambda$ is an arbitrary real constant. Eq.~\eqref{eigprob} is
an eigenvalue problem for $\Psi(x,s)$ with respect to the weight
$1/x$. The solutions that vanish at infinity take the form $
\Psi_\lambda(x,s) \propto e^{-\sqrt{2s} x}
U\left(-\lambda/\sqrt{2s},0,2 \sqrt{2s} x\right)$, but only for
$\lambda_k=\sqrt{2s} k, k=\{1,2,\dots\}$ do they vanish at $x=0$. The
correctly normalised eigenfunctions that satisfy boundary conditions
are therefore
\begin{equation}
\Psi_k(x,s)=\frac{ e^{-\sqrt{2s}\ x} U\left(-k,0,2 \sqrt{2s}  x\right)
}{\sqrt{k!(k-1)!}}
\end{equation}
These functions are an orthonormal set with respect to the weight
$1/x$, $\int_0^\infty dx \Psi_j(x,s)\Psi_k(x,s) \frac{1}{x} =
\delta_{j,k}$, and the corresponding closure relation reads
$\sum_{k=1}^\infty \Psi_k(x,s)\Psi_k(x',s)\frac{1}{x} =
\delta(x-x')$. One can use this to construct the solution of the
original equation. In particular
\begin{equation}
\label{zhat}
\hat{Z}(x,y,x_0,s) = \Theta(y)\sum_{k=1}^\infty \Psi_k(x,s)\Psi_k(x_0,s) e^{-\sqrt{2s} k y}
\end{equation}
We now return to the original problem of finding the probability of a
Brownian excursion with functional $\int_0^t  1/x(t') dt' = y(t)$. We make
use of the device $x_0=x=\epsilon$, and let $\epsilon\to0$ only after
normalization. In short,
\begin{equation}
\lim_{T\to\infty}
\text{Prob}(g_s(\Omega)/\sqrt{T}=y) = 
\left.
\lim_{\epsilon\to0} \frac{Z(\epsilon,\epsilon,y,t)}{Z_\epsilon}
\right|_{t=2}
\end{equation}
where $Z_\epsilon = \frac{1}{\sqrt{2\pi t}}(1-e^{-2\epsilon^2/t})$ is
the well-known normalising constant (see
e.g.~\cite{MajumdarComtet:2005}). From Eq.~\eqref{zhat} and expanding
for small $x=x_0=\epsilon$ term by term, we find
\begin{equation}
\frac{\hat{Z}(\epsilon,\epsilon,y,s)}{Z_\epsilon} \simeq
{\sqrt{2 \pi t}}\sum_{k=1}^\infty\frac{\Psi_k(\epsilon,s)^2}{(1-e^{-2 \epsilon^2/t})} e^{-\sqrt{2s} k y}
\end{equation}
Using the fact that $\Psi_k(\epsilon,s)^2 \simeq 8sk \epsilon^2$ for
small $\epsilon$, we finally arrive at
\begin{equation}
\label{limiting laplace}
\lim_{\epsilon \to 0}
\frac{\hat{Z}(\epsilon,\epsilon,y,s)}{Z_\epsilon} \simeq
\lim_{\epsilon \to 0}
{\sqrt{2 \pi t}}
\sum_{k=1}^\infty
\frac{8s k \epsilon^2}{2 \epsilon^2/t} e^{-\sqrt{2s} k y} =
{4s}{\sqrt{2 \pi}}t^{3/2}
\frac{e^{\sqrt{2s} y}}{\left(e^{\sqrt{2s}  y}-1\right)^2} =
{\sqrt{2 \pi}}t^{3/2}\frac{s}{\sinh^2( \sqrt{s/2} y)}
\end{equation}
Inverting terms involving $s$ yields
\begin{align}
\lim_{T\to\infty}
\text{Prob}(g_s(\Omega)/\sqrt{T}=y) &=
\left[
\frac{2\sqrt{2\pi}t^{3/2}\, \pi^2}{y^6} \sum_{k=1}^{\infty} (2k)^2
e^{-(2\pi k)^2 t /(2y^2)}((2\pi k)^2 t - 3y^2)
\right]_{t=2}
\\
 &=
\left[
\frac{2y}{t^2}\sum_{k=1}^\infty  e^{-(k y)^2/(2 t)}k^2 \left(k^2 y^2-3 t\right)
\right]_{t=2}
.
\end{align}
The first equation is obtained by collecting residues from double
poles, and is useful for a small $y$ expansion. The second equation
is obtained by expanding
Eq.~\eqref{limiting laplace} and inverting term by term, and is useful
for a large $y$ expansion. Both expressions converge rapidly and,
evaluating at $t=2$, are in excellent agreement with simulations,
see \Fref{limiting}.
\begin{figure}
\includegraphics[width=0.48\textwidth]{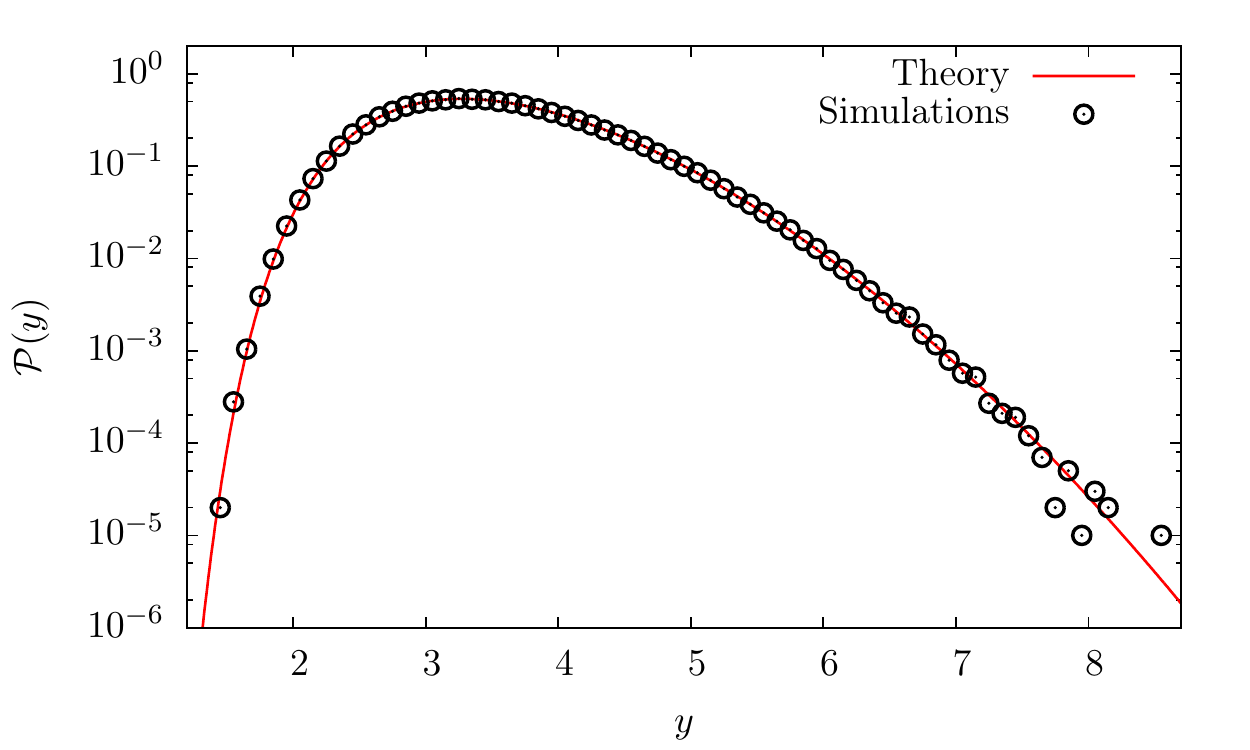}
\caption{\flabel{limiting} The distribution of $y=\int_0^2
  dt\ 1/x(t)$, where $x(t)$ is a Brownian excursion of length 2. The
  red full line is the analytical result and black symbols correspond
  to simulations.}
\end{figure}

\section{Laplace transform of $\pdf{g_s}{g_{s},h}$}\slabel{laplace_transform}

In this final appendix, we take yet another route in the calculation
of $\pdf{g_s}{g_{s},h}$ by finding its Laplace transform. The key
point in this approach is to approximate the embedded random walk of
the process by standard Brownian motion. Therefore, we expect our
approximation to hold as long as $T\gg1$. The approach is very similar
in spirit to that of \Aref{limiting_dist}, but both Appendices are
self-contained and can be read independently.\\

Let $x(t)$ denote the trajectory of a Brownian particle starting at
$x(0)=x_{0}$, and $t_{f}$ its first passage time to 0. Then we argue
that, in the Brownian motion picture, the original observable of
interest of the process $g_{s}$ corresponds to the quantity
$\mathcal{G}_{h}$,
\begin{equation}
\mathcal{G}_{h} = \int_{0}^{t_{f}} dt U_{h}(x(t)),
\end{equation}
with $U_{h}(x)=1/(x+h)$. Effectively, the underlying exponential
random variables $\xi(x(t))$ are replaced by their average. Such an
approximation, which can be seen as a self-averaging property of the
process, is well-justified because (i) the Brownian particle visits
any state infinitely many times, and (ii) the exponential distribution
has finite moments of any order. We are hence left with computing the
distribution of the integral of a function $U_h(x)$ along a Brownian
trajectory starting at $x(0)=x_{0}$ and ending at $x(t_{f})=0$. As
usual, the problem is most conveniently solved by taking the Laplace
transform of $\mathcal{G}_h$ (see the excellent review by Majumdar,
\cite{MajumdarComtet:2005}). In particular, the Laplace transform of
$\pdf{}{\mathcal{G}_h}$, which we denote by
$\hat{\mathcal{P}}(u;h,x_{0})$, fulfills the following differential
equation:
\begin{equation}
\frac{1}{2} \frac{\partial^2}{\partial x_0^2}\hat{\mathcal{P}}(u;h,x_{0})
-u\ U_h(x_0)\ \hat{\mathcal{P}}(u;h,x_{0})  = 0
\end{equation}
with boundary conditions $\lim_{x_{0}\to \infty}
\hat{\mathcal{P}}(u;h,x_{0}) = 0$ and $\lim_{x_{0}\to 0}
\hat{\mathcal{P}}(u;h,x_{0}) = 1$. Note that this is a differential
equation with respect to the initial position $x_{0}$. The general
solution to this differential equation is given by

\begin{equation}
\sqrt{2} C_1 \sqrt{u (h+x_0)} I_1\left(2 \sqrt{2} \sqrt{u (h+x_0)}\right)-\sqrt{2} C_2 \sqrt{u (h+x_0)} K_1\left(2 \sqrt{2} \sqrt{u (h+x_0)}\right)
\end{equation}
where $I_{1}(x)$ and $K_{1}(x)$ are modified Bessel functions of the
first and second kind respectively, and $C_{1}$ and $C_{2}$ are
constants to be determined via the boundary conditions. Because
$I_{1}(x_{0})$ diverges for $x_{0}\to \infty$, $C_1$ must be zero, and
$C_{2}$ is then fixed via the other boundary condition. Finally, by
setting $x_0=1$ we reach a remarkably simple expression for the
Laplace transform of $\pdf{g_s}{g_{s},h}$,
\begin{equation}\elabel{laplace_trans}
\hat{\mathcal{P}}(u;h) 
=\frac{\sqrt{u(h+1) }\ K_1\left(2 \sqrt{2} \sqrt{u(h+1) }\right)}{\sqrt{u h }\ K_1\left(2 \sqrt{2} \sqrt{u h}\right)}
\end{equation}
This result is not only of interest in itself, but also provides a
convenient way of evaluating $\pdf{g_s}{g_{s},h}$ by numerically
inverting \Eref{laplace_trans} (see \Fref{collapse} in the main
text). We can also recover the asymptotic exponents $\gamma_1,
\gamma_2$ of $\pdf{g_s}{g_{s},h}$ directly from its Laplace transform,
\Eref{laplace_trans}. To see this, we consider the first and second
derivatives of $\hat{\mathcal{P}}(u;h)$,
\begin{alignat}{2}
-\partial_u \hat{\mathcal{P}}(u;h) &\sim \sqrt{2/(hu)} &\text{ for } 1&\ll h \\
\partial_{uu}\hat{\mathcal{P}}(u;h) &\sim \frac{2}{u} &\text{ for } u&\ll1.
\end{alignat}
The first equation assumes large $h$, while the second does not; this
allows us to recover the two scaling regions mentioned in the main
text. Then it is easy to check that an application of a Tauberian theorem
\citep[p. 192]{Widder:1946} leads to \Eref{intermediate_regime} and
\Eref{large_regime} in the main text, recovering not only the
asymptotic exponents $\gamma_1, \gamma_2$, but also their associated
first order amplitudes.

\end{widetext}
\newcommand{\bibconferencename}[1]{\emph{#1}}
\bibliography{thresholding}

\begin{thebibliography}{28}
\expandafter\ifx\csname natexlab\endcsname\relax\def\natexlab#1{#1}\fi
\expandafter\ifx\csname bibnamefont\endcsname\relax
  \def\bibnamefont#1{#1}\fi
\expandafter\ifx\csname bibfnamefont\endcsname\relax
  \def\bibfnamefont#1{#1}\fi
\expandafter\ifx\csname citenamefont\endcsname\relax
  \def\citenamefont#1{#1}\fi
\expandafter\ifx\csname url\endcsname\relax
  \def\url#1{\texttt{#1}}\fi
\expandafter\ifx\csname urlprefix\endcsname\relax\def\urlprefix{URL }\fi
\providecommand{\bibinfo}[2]{#2}
\providecommand{\eprint}[2][]{\url{#2}}

\bibitem[{\citenamefont{Schorlemmer and
  Woessner}(2008)}]{schorlemmer2008probability}
\bibinfo{author}{\bibfnamefont{D.}~\bibnamefont{Schorlemmer}} \bibnamefont{and}
  \bibinfo{author}{\bibfnamefont{J.}~\bibnamefont{Woessner}},
  \bibinfo{journal}{Bulletin of the Seismological Society of America}
  \textbf{\bibinfo{volume}{98}}, \bibinfo{pages}{2103} (\bibinfo{year}{2008}).

\bibitem[{\citenamefont{Lovejoy et~al.}(2003)\citenamefont{Lovejoy, Lilley,
  Desaulniers-Soucy, and Schertzer}}]{lovejoy2003large}
\bibinfo{author}{\bibfnamefont{S.}~\bibnamefont{Lovejoy}},
  \bibinfo{author}{\bibfnamefont{M.}~\bibnamefont{Lilley}},
  \bibinfo{author}{\bibfnamefont{N.}~\bibnamefont{Desaulniers-Soucy}},
  \bibnamefont{and}
  \bibinfo{author}{\bibfnamefont{D.}~\bibnamefont{Schertzer}},
  \bibinfo{journal}{Physical Review E} \textbf{\bibinfo{volume}{68}},
  \bibinfo{pages}{025301} (\bibinfo{year}{2003}).

\bibitem[{\citenamefont{Paczuski et~al.}(2005)\citenamefont{Paczuski,
  Boettcher, and Baiesi}}]{Paczuski_btw}
\bibinfo{author}{\bibfnamefont{M.}~\bibnamefont{Paczuski}},
  \bibinfo{author}{\bibfnamefont{S.}~\bibnamefont{Boettcher}},
  \bibnamefont{and} \bibinfo{author}{\bibfnamefont{M.}~\bibnamefont{Baiesi}},
  \bibinfo{journal}{Phys. Rev. Lett.} \textbf{\bibinfo{volume}{95}},
  \bibinfo{pages}{181102} (\bibinfo{year}{2005}).

\bibitem[{\citenamefont{Bak and Sneppen}(1993)}]{BakSneppen:1993}
\bibinfo{author}{\bibfnamefont{P.}~\bibnamefont{Bak}} \bibnamefont{and}
  \bibinfo{author}{\bibfnamefont{K.}~\bibnamefont{Sneppen}},
  \bibinfo{journal}{Phys. Rev. Lett.} \textbf{\bibinfo{volume}{71}},
  \bibinfo{pages}{4083} (\bibinfo{year}{1993}).

\bibitem[{\citenamefont{Pruessner}(2012)}]{Pruessner:2012:Book}
\bibinfo{author}{\bibfnamefont{G.}~\bibnamefont{Pruessner}},
  \emph{\bibinfo{title}{Self-Organised Criticality}}
  (\bibinfo{publisher}{Cambridge University Press},
  \bibinfo{address}{Cambridge, UK}, \bibinfo{year}{2012}).

\bibitem[{\citenamefont{Paczuski et~al.}(1996)\citenamefont{Paczuski, Maslov,
  and Bak}}]{PaczuskiMaslovBak:1996}
\bibinfo{author}{\bibfnamefont{M.}~\bibnamefont{Paczuski}},
  \bibinfo{author}{\bibfnamefont{S.}~\bibnamefont{Maslov}}, \bibnamefont{and}
  \bibinfo{author}{\bibfnamefont{P.}~\bibnamefont{Bak}},
  \bibinfo{journal}{Phys. Rev. E} \textbf{\bibinfo{volume}{53}},
  \bibinfo{pages}{414} (\bibinfo{year}{1996}), \eprint{arXiv:adap-org/9510002}.

\bibitem[{\citenamefont{Sneppen}(1995)}]{Sneppen:1995b}
\bibinfo{author}{\bibfnamefont{K.}~\bibnamefont{Sneppen}}, in
  \emph{\bibinfo{booktitle}{Scale Invariance, Interfaces, and Non-Equilibrium
  Dynamics}}, edited by
  \bibinfo{editor}{\bibfnamefont{A.}~\bibnamefont{McKane}},
  \bibinfo{editor}{\bibfnamefont{M.}~\bibnamefont{Droz}},
  \bibinfo{editor}{\bibfnamefont{J.}~\bibnamefont{Vannimenus}},
  \bibnamefont{and} \bibinfo{editor}{\bibfnamefont{D.}~\bibnamefont{Wolf}}
  (\bibinfo{publisher}{Plenum Press}, \bibinfo{address}{New York, NY, USA},
  \bibinfo{year}{1995}), pp. \bibinfo{pages}{295--302}, \bibinfo{note}{{NATO}
  Advanced Study Institute on \bibconferencename{Scale Invariance, Interfaces,
  and Non-Equilibrium Dynamics}, Cambridge, UK, Jun 20--30, 1994}.

\bibitem[{\citenamefont{Grassberger}(1995)}]{Grassberger:1995}
\bibinfo{author}{\bibfnamefont{P.}~\bibnamefont{Grassberger}},
  \bibinfo{journal}{Phys. Lett. A} \textbf{\bibinfo{volume}{200}},
  \bibinfo{pages}{277} (\bibinfo{year}{1995}).

\bibitem[{\citenamefont{Garber et~al.}(2009)\citenamefont{Garber, Hallerberg,
  and Kantz}}]{Garber_pre}
\bibinfo{author}{\bibfnamefont{A.}~\bibnamefont{Garber}},
  \bibinfo{author}{\bibfnamefont{S.}~\bibnamefont{Hallerberg}},
  \bibnamefont{and} \bibinfo{author}{\bibfnamefont{H.}~\bibnamefont{Kantz}},
  \bibinfo{journal}{Phys. Rev. E} \textbf{\bibinfo{volume}{80}},
  \bibinfo{pages}{026124} (\bibinfo{year}{2009}).

\bibitem[{\citenamefont{Gardiner}(1997)}]{Gardiner:1997}
\bibinfo{author}{\bibfnamefont{C.~W.} \bibnamefont{Gardiner}},
  \emph{\bibinfo{title}{Handbook of Stochastic Methods}}
  (\bibinfo{publisher}{Springer-Verlag}, \bibinfo{address}{Berlin, Germany},
  \bibinfo{year}{1997}), \bibinfo{edition}{2nd} ed.

\bibitem[{\citenamefont{Hinrichsen}(2000)}]{Hinrichsen:2000a}
\bibinfo{author}{\bibfnamefont{H.}~\bibnamefont{Hinrichsen}},
  \bibinfo{journal}{Adv. Phys.} \textbf{\bibinfo{volume}{49}},
  \bibinfo{pages}{815} (\bibinfo{year}{2000}),
  \eprint{arXiv:cond-mat/0001070v2}.

\bibitem[{\citenamefont{Harris}(1963)}]{Harris:1963}
\bibinfo{author}{\bibfnamefont{T.~E.} \bibnamefont{Harris}},
  \emph{\bibinfo{title}{The Theory of Branching Processes}}
  (\bibinfo{publisher}{Springer-Verlag}, \bibinfo{address}{Berlin, Germany},
  \bibinfo{year}{1963}).

\bibitem[{\citenamefont{Peters et~al.}(2002)\citenamefont{Peters, Hertlein, and
  Christensen}}]{PetersHertleinChristensen:2002}
\bibinfo{author}{\bibfnamefont{O.}~\bibnamefont{Peters}},
  \bibinfo{author}{\bibfnamefont{C.}~\bibnamefont{Hertlein}}, \bibnamefont{and}
  \bibinfo{author}{\bibfnamefont{K.}~\bibnamefont{Christensen}},
  \bibinfo{journal}{Phys. Rev. Lett.} \textbf{\bibinfo{volume}{88}},
  \bibinfo{eid}{018701} (pages~\bibinfo{numpages}{4}) (\bibinfo{year}{2002}),
  \eprint{arXiv:cond-mat/0201468}.

\bibitem[{\citenamefont{Galassi et~al.}(2009)\citenamefont{Galassi, Davies,
  Theiler, Gough, Jungman, Alken, Booth, and Rossi}}]{GalassiETAL:2009}
\bibinfo{author}{\bibfnamefont{M.}~\bibnamefont{Galassi}},
  \bibinfo{author}{\bibfnamefont{J.}~\bibnamefont{Davies}},
  \bibinfo{author}{\bibfnamefont{J.}~\bibnamefont{Theiler}},
  \bibinfo{author}{\bibfnamefont{B.}~\bibnamefont{Gough}},
  \bibinfo{author}{\bibfnamefont{G.}~\bibnamefont{Jungman}},
  \bibinfo{author}{\bibfnamefont{P.}~\bibnamefont{Alken}},
  \bibinfo{author}{\bibfnamefont{M.}~\bibnamefont{Booth}}, \bibnamefont{and}
  \bibinfo{author}{\bibfnamefont{F.}~\bibnamefont{Rossi}},
  \emph{\bibinfo{title}{GNU Scientific Library Reference Manual}}
  (\bibinfo{publisher}{Network Theory Ltd.}, \bibinfo{year}{2009}),
  \bibinfo{edition}{3rd} ed.,
  \bibinfo{note}{\url{http://www.network-theory.co.uk/gsl/manual/}, accessed 18
  Aug 2009}.

\bibitem[{\citenamefont{Deluca and Corral}(2013)}]{DelucaCorral}
\bibinfo{author}{\bibfnamefont{A.}~\bibnamefont{Deluca}} \bibnamefont{and}
  \bibinfo{author}{\bibfnamefont{A.}~\bibnamefont{Corral}},
  \bibinfo{journal}{Acta Geophys.} \textbf{\bibinfo{volume}{61}}
  (\bibinfo{year}{2013}).

\bibitem[{\citenamefont{Rubin et~al.}(2014)\citenamefont{Rubin, Pruessner, and
  Pavliotis}}]{RubinPruessnerPavliotis:2014}
\bibinfo{author}{\bibfnamefont{K.~J.} \bibnamefont{Rubin}},
  \bibinfo{author}{\bibfnamefont{G.}~\bibnamefont{Pruessner}},
  \bibnamefont{and} \bibinfo{author}{\bibfnamefont{G.~A.}
  \bibnamefont{Pavliotis}}, \bibinfo{journal}{J. Phys. A}
  \textbf{\bibinfo{volume}{47}}, \bibinfo{pages}{195001}
  (\bibinfo{year}{2014}), \eprint{arXiv:1401.0695}.

\bibitem[{\citenamefont{Majumdar and Comtet}(2004)}]{MajumdarComtet:2004}
\bibinfo{author}{\bibfnamefont{S.~N.} \bibnamefont{Majumdar}} \bibnamefont{and}
  \bibinfo{author}{\bibfnamefont{A.}~\bibnamefont{Comtet}},
  \bibinfo{journal}{Phys. Rev. Lett.} \textbf{\bibinfo{volume}{92}},
  \bibinfo{eid}{225501} (pages~\bibinfo{numpages}{4}) (\bibinfo{year}{2004}).

\bibitem[{\citenamefont{Majumdar and Comtet}(2005)}]{MajumdarComtet:2005}
\bibinfo{author}{\bibfnamefont{S.~N.} \bibnamefont{Majumdar}} \bibnamefont{and}
  \bibinfo{author}{\bibfnamefont{A.}~\bibnamefont{Comtet}},
  \bibinfo{journal}{J. Stat. Phys.} \textbf{\bibinfo{volume}{119}},
  \bibinfo{pages}{777} (\bibinfo{year}{2005}).

\bibitem[{\citenamefont{Mohanty}(1979)}]{Mohanty:1979}
\bibinfo{author}{\bibfnamefont{G.}~\bibnamefont{Mohanty}},
  \emph{\bibinfo{title}{Lattice path counting and applications}}
  (\bibinfo{publisher}{Academic Press New York}, \bibinfo{year}{1979}).

\bibitem[{\citenamefont{Christensen et~al.}(2008)\citenamefont{Christensen,
  Farid, Pruessner, and Stapleton}}]{ChristensenETAL:2008}
\bibinfo{author}{\bibfnamefont{K.}~\bibnamefont{Christensen}},
  \bibinfo{author}{\bibfnamefont{N.}~\bibnamefont{Farid}},
  \bibinfo{author}{\bibfnamefont{G.}~\bibnamefont{Pruessner}},
  \bibnamefont{and}
  \bibinfo{author}{\bibfnamefont{M.}~\bibnamefont{Stapleton}},
  \bibinfo{journal}{Eur. Phys. J. B} \textbf{\bibinfo{volume}{62}},
  \bibinfo{pages}{331} (\bibinfo{year}{2008}).

\bibitem[{\citenamefont{Corral}(2009)}]{Corral_jstat}
\bibinfo{author}{\bibfnamefont{A.}~\bibnamefont{Corral}}, \bibinfo{journal}{J.
  Stat. Mech.} \textbf{\bibinfo{volume}{P01022}} (\bibinfo{year}{2009}).

\bibitem[{\citenamefont{Laurson et~al.}(2009)\citenamefont{Laurson, Illa, and
  Alava}}]{Laurson_upon}
\bibinfo{author}{\bibfnamefont{L.}~\bibnamefont{Laurson}},
  \bibinfo{author}{\bibfnamefont{X.}~\bibnamefont{Illa}}, \bibnamefont{and}
  \bibinfo{author}{\bibfnamefont{M.~J.} \bibnamefont{Alava}},
  \bibinfo{journal}{Journal of Statistical Mechanics: Theory and Experiment}
  \textbf{\bibinfo{volume}{2009}}, \bibinfo{pages}{P01019}
  (\bibinfo{year}{2009}), \eprint{arXiv:0810.0948}.

\bibitem[{\citenamefont{Knuth}(1997)}]{Knuth:1997V1}
\bibinfo{author}{\bibfnamefont{D.~E.} \bibnamefont{Knuth}},
  \emph{\bibinfo{title}{Fundamental Algorithms}}, vol.~\bibinfo{volume}{1} of
  \emph{\bibinfo{series}{The Art of Computer Programming}}
  (\bibinfo{publisher}{Addison-Wesley}, \bibinfo{address}{Reading, MA, USA},
  \bibinfo{year}{1997}), \bibinfo{edition}{3rd} ed.

\bibitem[{\citenamefont{Stanley}(1999)}]{Stanley:1999}
\bibinfo{author}{\bibfnamefont{R.~P.} \bibnamefont{Stanley}},
  \emph{\bibinfo{title}{Enumerative Combinatorics, Volume II}},
  no.~\bibinfo{number}{62} in \bibinfo{series}{Cambridge Studies in Advanced
  Mathematics} (\bibinfo{publisher}{Cambridge University Press},
  \bibinfo{address}{Cambridge, UK}, \bibinfo{year}{1999}).

\bibitem[{\citenamefont{{Wolfram Research Inc.}}(2011)}]{Mathematica:8.0.1.0}
\bibinfo{author}{\bibnamefont{{Wolfram Research Inc.}}},
  \emph{\bibinfo{title}{Mathematica}} (\bibinfo{publisher}{Wolfram Research,
  Inc.}, \bibinfo{address}{Champaign, IL, USA}, \bibinfo{year}{2011}),
  \bibinfo{note}{version 8.0.1.0}.

\bibitem[{\citenamefont{Abramowitz and Stegun}(1970)}]{AbramowitzStegun:1970}
\bibinfo{editor}{\bibfnamefont{M.}~\bibnamefont{Abramowitz}} \bibnamefont{and}
  \bibinfo{editor}{\bibfnamefont{I.~A.} \bibnamefont{Stegun}}, eds.,
  \emph{\bibinfo{title}{Handbook of Mathematical Functions}}
  (\bibinfo{publisher}{Dover Publications, Inc.}, \bibinfo{address}{New York,
  NY, USA}, \bibinfo{year}{1970}).

\bibitem[{\citenamefont{Chaichian and Demichev}(2001)}]{Chaichian:2001}
\bibinfo{author}{\bibfnamefont{M.}~\bibnamefont{Chaichian}} \bibnamefont{and}
  \bibinfo{author}{\bibfnamefont{A.}~\bibnamefont{Demichev}},
  \emph{\bibinfo{title}{Path Integrals in Physics: Volume I Stochastic
  Processes and Quantum Mechanics}}, Institute of physics series in
  mathematical and computational physics (\bibinfo{publisher}{Taylor \&
  Francis}, \bibinfo{year}{2001}), ISBN \bibinfo{isbn}{9780750308014}.

\bibitem[{\citenamefont{Widder}(1946)}]{Widder:1946}
\bibinfo{author}{\bibfnamefont{D.~V.} \bibnamefont{Widder}},
  \emph{\bibinfo{title}{The Laplace Transform}} (\bibinfo{publisher}{Princeton
  University Press}, \bibinfo{address}{Princeton}, \bibinfo{year}{1946}).

\end{thebibliography}
\end{document}